\documentclass[12pt]{article}
\pdfoutput=1
\usepackage[colorlinks,linkcolor=Blue,citecolor=Blue,bookmarks,bookmarksnumbered]{hyperref}
\usepackage[scaled=0.85]{helvet}
\usepackage{amsmath,amssymb,accents,mathrsfs,XoohmE}
\usepackage{graphicx,color}
\usepackage{booktabs}
\usepackage{multirow}
\usepackage{placeins}
\usepackage{amsmath}

\usepackage{XoohmE}

\definecolor{Green}  {rgb}{0.10,0.70,0.10} 
\definecolor{Orange} {rgb}{1.00,0.50,0.15} 
\definecolor{Red}    {rgb}{0.90,0.00,0.12} 
\definecolor{Purple} {rgb}{0.50,0.25,0.55} 
\definecolor{Turque} {rgb}{0.00,0.65,0.85} 
\definecolor{Blue}   {rgb}{0.00,0.00,1.00} 
\definecolor{Magenta}{rgb}{1.00,0.00,1.00} 
\definecolor{Gold}   {rgb}{1.00,0.75,0.25} 
\definecolor{Seaweed}{rgb}{0.01,0.24,0.09} 
\definecolor{Brown}  {rgb}{0.43,0.26,0.32} 
\definecolor{grey1}  {rgb}{0.20,0.20,0.20} 
\definecolor{grey2}  {rgb}{0.40,0.40,0.40} 
\definecolor{grey3}  {rgb}{0.60,0.60,0.60} 
\definecolor{grey4}  {rgb}{0.80,0.80,0.80} 
\definecolor{grey5}  {rgb}{0.90,0.90,0.90} 
\def\C#1#2{{\ifcase#1\or
             \color{Green}\or \color{Orange}\or \color{Red}\or
              \color{Purple}\or \color{Turque}\or \color{Blue}\or
               \color{Magenta}\or \color{Gold}\or \color{Seaweed}\or
                \color{Brown}\or\color{grey1}\or\color{grey2}\or
                 \color{grey3}\else\color{grey4}\fi#2}}

\definecolor{Slate} {rgb}{0.00,0.45,0.55}

\def\fracm#1#2{\hbox{\large{${\frac{{#1}}{{#2}}}$}}}

\def\be{\begin{equation}}
\def\ee{\end{equation}}
\newcommand{\bea}{\begin{eqnarray}}
\newcommand{\eea}{\end{eqnarray}}
\newcommand{\ena}{\end{eqnarray}}

\def\pp{{\mathchoice
          {
              \kern 1pt%
              \raise 1pt
              \vbox{\hrule width5pt height0.4pt depth0pt
                    \kern -2pt
                    \hbox{\kern 2.3pt
                          \vrule width0.4pt height6pt depth0pt
                          }
                    \kern -2pt
                    \hrule width5pt height0.4pt depth0pt}%
                    \kern 1pt
           }
            {
              \kern 1pt%
              \raise 1pt
              \vbox{\hrule width4.3pt height0.4pt depth0pt
                    \kern -1.8pt
                    \hbox{\kern 1.95pt
                          \vrule width0.4pt height5.4pt depth0pt
                          }
                    \kern -1.8pt
                    \hrule width4.3pt height0.4pt depth0pt}%
                    \kern 1pt
            }
            {
              \kern 0.5pt%
              \raise 1pt
              \vbox{\hrule width4.0pt height0.3pt depth0pt
                    \kern -1.9pt  
                    \hbox{\kern 1.85pt
                          \vrule width0.3pt height5.7pt depth0pt
                          }
                    \kern -1.9pt
                    \hrule width4.0pt height0.3pt depth0pt}%
                    \kern 0.5pt
            }
            {
              \kern 0.5pt%
              \raise 1pt
              \vbox{\hrule width3.6pt height0.3pt depth0pt
                    \kern -1.5pt
                    \hbox{\kern 1.65pt
                          \vrule width0.3pt height4.5pt depth0pt
                          }
                    \kern -1.5pt
                    \hrule width3.6pt height0.3pt depth0pt}%
                    \kern 0.5pt
            }
        }}

\def\mm{{\mathchoice
                       {
                             \kern 1pt
               \raise 1pt    \vbox{\hrule width5pt height0.4pt depth0pt
                                  \kern 2pt
                                  \hrule width5pt height0.4pt depth0pt}
                             \kern 1pt}
                       {
                            \kern 1pt
               \raise 1pt \vbox{\hrule width4.3pt height0.4pt depth0pt
                                  \kern 1.8pt
                                  \hrule width4.3pt height0.4pt depth0pt}
                             \kern 1pt}
                       {
                            \kern 0.5pt
               \raise 1pt
                            \vbox{\hrule width4.0pt height0.3pt depth0pt
                                  \kern 1.9pt
                                  \hrule width4.0pt height0.3pt depth0pt}
                            \kern 1pt}
                       {
                           \kern 0.5pt
             \raise 1pt  \vbox{\hrule width3.6pt height0.3pt depth0pt
                                  \kern 1.5pt
                                  \hrule width3.6pt height0.3pt depth0pt}
                           \kern 0.5pt}
                       }}

\def\ad{{\kern0.5pt
                   \alpha \kern-5.05pt \raise5.8pt\hbox{$\textstyle.$}\kern
0.5pt}}

\def\bd{{\kern0.5pt
                   \beta \kern-5.05pt \raise5.8pt\hbox{$\textstyle.$}\kern
0.5pt}}

\def\qd{{\kern0.5pt
                   q \kern-5.05pt \raise5.8pt\hbox{$\textstyle.$}\kern
0.5pt}}
\def\Dot#1{{\kern0.5pt
     {#1} \kern-5.05pt \raise5.8pt\hbox{$\textstyle.$}\kern
0.5pt}}

\catcode`@=11
\def\un#1{\relax\ifmmode\@@underline#1\else
        $\@@underline{\hbox{#1}}$\relax\fi}
\catcode`@=12

\def\a{\alpha}
\def\b{\beta}
\def\c{\chi}
\def\d{\delta}
\def\e{\epsilon}

\def\g{\gamma}

\def\i{\iota}
\def\j{\psi}
\def\k{\kappa}
\def\l{\lambda}

\def\o{\omega}

\def\r{\rho}
\def\s{\sigma}
\def\t{\tau}

\def\F{\Phi}

\def\L{\Lambda}
\def\O{\Omega}
\def\P{\Pi}

\def\S{\Sigma}

\def\ca{{\cal A}}

\def\cf{{\cal F}}

\def\dslash{\not{\hbox{\kern-2pt $\partial$}}}
\def\Dslash{\not{\hbox{\kern-4pt $D$}}}
\def\pslash{\not{\hbox{\kern-2.3pt $p$}}}
 \newtoks\slashfraction
 \slashfraction={.13}
 \def\slash#1{\setbox0\hbox{$ #1 $}
 \setbox0\hbox to \the\slashfraction\wd0{\hss \box0}/\box0 }
 
\def\Sc#1{{\hbox{\sc #1}}}      
\def\kcr{{\hbox{\ro \char'170}}}                
\def\ktl{{\hbox{\ro \char'170}}}        
\def\ktr{{\hbox{\ro \char'170}}}        
\def\kbl{{\hbox{\ro \char'170}}}        
\def\kbr{{\hbox{\ro \char'170}}}        

\def\plpl{\raise-2pt\hbox{$\raise3pt\hbox{$_+$}\hskip-6.67pt\raise0.0pt
\hbox{$^+$}\hskip 0.01pt$}}
\def\mimi{\raise-2pt\hbox{$\raise3pt\hbox{$_-$}\hskip-6.67pt\raise0.0pt
\hbox{$^-$}\hskip 0.01pt$}} 

\def\bo{{\raise.15ex\hbox{\large$\Box$}}}               
\def\pa{\partial}                                       
\def\de{\nabla}                                         
\def\TH{{\raise.2ex\hbox{$\displaystyle \bigodot$}\mskip-4.7mu \llap H \;}}
\def\face{{\raise.2ex\hbox{$\displaystyle \bigodot$}\mskip-2.2mu \llap {$\ddot
        \smile$}}}                                      

\def\dt#1{\on{\hbox{\bf .}}{#1}}                
\def\Dot#1{\dt{#1}}

\def\Bar#1{\overline{#1}}                       
\def\leftrightarrowfill{$\mathsurround=0pt \mathord\leftarrow \mkern-6mu
        \cleaders\hbox{$\mkern-2mu \mathord- \mkern-2mu$}\hfill
        \mkern-6mu \mathord\rightarrow$}
\def\dvec#1{\vbox{\ialign{##\crcr
        \leftrightarrowfill\crcr\noalign{\kern-1pt\nointerlineskip}
        $\hfil\displaystyle{#1}\hfil$\crcr}}}           
\def\dt#1{{\buildrel {\hbox{\LARGE .}} \over {#1}}}     

\def\fracm#1#2{\hbox{\large{${\frac{{#1}}{{#2}}}$}}}
\def\sfrac#1#2{{\vphantom1\smash{\lower.5ex\hbox{\small$#1$}}\over
        \vphantom1\smash{\raise.4ex\hbox{\small$#2$}}}} 
\def\bfrac#1#2{{\vphantom1\smash{\lower.5ex\hbox{$#1$}}\over
        \vphantom1\smash{\raise.3ex\hbox{$#2$}}}}       
\def\afrac#1#2{{\vphantom1\smash{\lower.5ex\hbox{$#1$}}\over#2}}    

\def\pa{\partial}

\def\ad{{\Dot{\alpha}}}
\def\bd{{\Dot{\beta}}}

 \font\rOpe=cmsy10                        
 \def\ktl{{\hbox{\rOpe\char'170}}}        
 \def\kbl{{\hbox{\rOpe\char'170}}}        
 \def\kcr{{\reflectbox{\rOpe\char'170}}}        
 \def\ktr{{\reflectbox{\rOpe\char'170}}}        
 \def\kbr{{\reflectbox{\rOpe\char'170}}}        
 \def\Border{\vbox{\hsize0pt
        \setlength{\unitlength}{1mm}
        \newcount\xco
        \newcount\yco
        \xco=-21
        \yco=12
        \begin{picture}(0,0)(-7.5,0)
        \put(\xco,\yco){$\ktl$}
        \advance\yco by-1
        {\loop
        \put(\xco,\yco){$\kcr$}
        \advance\yco by-2
        \ifnum\yco>-240
        \repeat
        \put(\xco,\yco){$\kbl$}}
        \xco=170
        \yco=12
        \put(\xco,\yco){$\ktr$}
        \advance\yco by-1
        {\loop
        \put(\xco,\yco){$\kcr$}
        \advance\yco by-2
        \ifnum\yco>-240
        \repeat
        \put(\xco,\yco){$\kbr$}}
        \put(-19.5,13){\scalebox{.6065}{%
         University of Maryland Center for String and Particle  Theory \&\ Physics Department%
        |University of Maryland Center for String and Particle  Theory \&\ Physics Department}}
        \put(-19.5,-241.5){\scalebox{.5835}{%
         ****University of Maryland * Center for String and
         Particle  Theory* Physics Department****University of Maryland *Center
        for String and Particle  Theory* Physics Department}}
        \end{picture}
        \par\vskip-8mm}}
\definecolor{UMred}{rgb}{.9,.05,.2}
\definecolor{HUblue}{rgb}{.0,.3,.7}

\definecolor{Red}    {rgb}{0.90,0.00,0.12} 
\definecolor{Blue}   {rgb}{0.00,0.00,1.00} 
\definecolor{Green}  {rgb}{0.10,0.70,0.10} 
\definecolor{Turque} {rgb}{0.00,0.65,0.85} 
\definecolor{Orange} {rgb}{1.00,0.50,0.15} 
\definecolor{Magenta}{rgb}{1.00,0.00,1.00} 
\definecolor{Gold}   {rgb}{1.00,0.75,0.25} 
\definecolor{Seaweed}{rgb}{0.01,0.24,0.09} 
\definecolor{Purple} {rgb}{0.50,0.25,0.55} 
\definecolor{Brown}  {rgb}{0.43,0.26,0.32} 
\definecolor{grey1}  {rgb}{0.20,0.20,0.20} 
\definecolor{grey2}  {rgb}{0.40,0.40,0.40} 
\definecolor{grey3}  {rgb}{0.60,0.60,0.60} 
\definecolor{grey4}  {rgb}{0.80,0.80,0.80} 
\definecolor{grey5}  {rgb}{0.90,0.90,0.90} 
\def\C#1#2{{\ifcase#1\or
             \color{Red}\or \color{Green}\or \color{Blue}\or\
              \color{Turque}\or \color{Orange}\or \color{Magenta}\or 
               \color{Gold}\or \color{Seaweed}\or \color{Purple}\or
                \color{Brown}\or\color{grey1}\or\color{grey2}\or
                 \color{grey3}\else\color{grey4}\fi#2}}

\definecolor{Slate} {rgb}{0.00,0.45,0.55}

\newdimen\parshift\parshift=\parindent
\catcode`@=11
 \long\def\@footnotetext#1{\insert\footins{\reset@font\footnotesize
           \interlinepenalty\interfootnotelinepenalty\splittopskip%
            \footnotesep\splitmaxdepth\dp\strutbox\floatingpenalty\@MM%
             \hsize\columnwidth\addtolength{\hsize}{-2\parindent}
              \@parboxrestore\protected@edef\@currentlabel%
              {\csname p@footnote\endcsname\@thefnmark}%
                \color@begingroup%
                 \@makefntext{\rule\z@\footnotesep\ignorespaces#1%
                  \@finalstrut\strutbox}%
                \color@endgroup}}
 \long\def\@makefntext#1{\hglue\parshift%
           \vbox{\noindent\baselineskip=11pt plus.5pt minus.5pt\hb@xt@0em{\hss\@makefnmark\kern1pt}#1}}
\catcode`@=12

\newskip\humongous \humongous=0pt plus 1000pt minus 1000pt
\def\caja{\mathsurround=0pt}
\def\eqalign#1{\,\vcenter{\openup2\jot \caja
        \ialign{\strut \hfil$\displaystyle{##}$&$
        \displaystyle{{}##}$\hfil\crcr#1\crcr}}\,}
\newif\ifdtup

\makeatletter
\def\section{\@startsection{section}{1}{\z@}
        {3ex plus-1ex minus-.2ex}{1pt plus1pt}{\large\sf\bfseries\boldmath}}
\def\subsection{\@startsection{subsection}{2}{\z@}
         {1.5ex plus-1ex minus-.2ex}{0.01pt plus1pt}{\sf\slshape}}
\def\subsubsection{\@startsection{subsubsection}{3}{\z@}
          {1.5ex plus-1ex minus-.2ex}{0.01pt plus0.2pt}{\sf\boldmath}}
\def\paragraph{\@startsection{paragraph}{4}{\z@}
           {.75ex \@plus.5ex \@minus.2ex}{-2mm}{\sf\bfseries\boldmath}}
\makeatother

 \allowdisplaybreaks
 \seceq

\begin{document}

\thispagestyle{empty}
%
\noindent{\small
\hfill{HET-1779}  \\ 
$~~~~~~~~~~~~~~~~~~~~~~~~~~~~~~~~~~~~~~~~~~~~~~~~~~~~~~~~~~~~~~~~~$
$~~~~~~~~~~~~~~~~~~~~~~~~~~~~~~~~~~~~~~~~~~~~~~~~~~~~~~~~~~~~~~~~~$
{}
}
\vspace*{8mm}
\begin{center}
{\large \bf
On Linearized Nordstr\" om Supergravity in \vskip0.1pt
Eleven and Ten Dimensional Superspaces}   \\   [12mm]
{\large {
S.\ James Gates, Jr.,\footnote{sylvester$_-$gates@brown.edu}$^{a}$}
Yangrui Hu\footnote{yangrui$_-$hu@brown.edu}$^{a}$,
Hanzhi Jiang\footnote{hanzhi$_-$jiang@alumni.brown.edu}$^{a,b}$, 
and S.-N. Hazel Mak\footnote{sze$_-$ning$_-$mak@brown.edu}$^{a}$
}
\\*[12mm]
\emph{
\centering
$^{a}$Department of Physics, Brown University,
\\[1pt]
Box 1843, 182 Hope Street, Barus \& Holley 545,
Providence, RI 02912, USA 
\\[12pt]
${}^{b}$Department of Physics \& Astronomy,
\\[1pt]
Rutgers University,
 Piscataway, NJ 08855-0849, USA
}
 \\*[78mm]
{ ABSTRACT}\\[4mm]
\parbox{142mm}{\parindent=2pc\indent\baselineskip=14pt plus1pt
As the full off-shell theories of supergravity in 
the important dimensions of eleven and ten dimensions are currently 
unknown, we introduce a superfield formalism as a foundation and 
experimental laboratory to explore the possibility that the scalar 
versions of the higher dimensional supergravitation theory can be 
constructed.}
 \end{center}
\vfill
\vfill
\clearpage
%

\newpage
\section{Introduction}

In 1907, Einstein described his ``happiest thought'' \cite{Einst1} which marked the
commencement of the race to create the Theory of General Relativity.  Unrealized,
he was already decidedly at a disadvantage.  As early 1900 the astronomer Karl 
Schwarzschild (1873-1916) \cite{SchWrz} had written about Riemann's geometrical 
concepts to describe curved space - but not curved space-time.  The latter would 
not emerge until Hermann Minkowski introduced the concept of ``four-geometry"
into physics \cite{Mink1,Mink2}.

By 1914, there were a number of competitors.  At a minimum, these included 
Max Abraham (1875-1922), Gustav Mie (1868-1957), and Gunnar Nordstr\" om 
(1881-1923).  Even the accomplished mathematician David Hilbert (1862-1943) 
became involved but at the conclusion made the comment in his own work,
``The differential equations of gravitation that result are, as it seems to me, in 
 agreement with the magnificent theory of general relativity established by Einstein 
 \cite{Einst2}.''

Along the pathway to the end of the race, the idea of scalar gravitational theories 
was explored.  It is of note that Nordstr\" om created two such sets of equations 
\cite{N1,N2} and even Einstein was taken with idea before discarding it.  In the 
scalar approach, the usual metric of the space-time manifold is replaced by a 
single scalar field.  A way to do this is to begin with the Minkowski metric and simply 
multiply it by a scalar field.  This implies that, geometrically, scalar theories of 
gravitation are all members of the same conformal class as the usual flat Minkowski 
metric.  Mathematically, scalar gravitation theories are perfectly consistent... they 
simply do not describe the physical laws observed in our universe.  

In the nineteen eighties, superspace geometrical descriptions of supergravity 
in eleven and ten dimensions were presented in the physics literature.  To the best of
our knowledge a list of these inaugural publications looks as:
\newline \noindent
$~~~~~$ $~~~~~$ (a.)
11D, $\cal N$ = 1 supergravity \cite{crD11a,crD11b},
\newline \noindent
$~~~~~$ $~~~~~$ (b.)
10D, $\cal N$ = 2A supergravity, \cite{crD10A},
\newline \noindent
$~~~~~$ $~~~~~$ (c.)
10D, $\cal N$ = 2B supergravity \cite{crD10B1,crD10B2}, and
\newline \noindent
$~~~~~$ $~~~~~$ (d.)
10D, $\cal N$ = 1 supergravity \cite{crD10N}.
\newline \noindent
Of course these theories had been obtained in other descriptions even earlier.  Interested
parties are directed to these works for such references.

If we think of the references in \cite{crD11a,crD11b,crD10A,crD10B1,crD10B2,crD10N} as the 
eleven and ten dimensional analogs of Einstein's ``happiest thought,'' then by 
analogy all that have occurred since these works are analogs of the race to find the
Theory of General Relativity.  This also reveals a glaring disappointment.  Since the 
``bell was rung'' in this new race, all the competition is still at the starting line.

How would one know the race has been successfully ended?  As indicated by the
title of the work \cite{crD11b} (``Eleven-Dimensional Supergravity on the Mass-Shell 
in Superspace''), these descriptions possess sets of Bianchi identities that are consistent
{\em {only}} when the equations of motion for the component fields in the theory 
satisfy their mass-shell conditions.  This holds true for all of the works in 
\cite{crD11a,crD11b,crD10A,crD10B1,crD10B2,crD10N}.  So we may take as the sign of the 
successful completion of the race, if a set of superspace geometries were explicitly 
found such that their Bianchi identities do {\em {not}} require a mass shell condition.
By way of comparison, for 4D, $\cal N$ = 1 superspace supergravity, the analogs of 
the ``happiest thought'' and the conclusion of the race occurred within one year as
is seen via works completed by Wess and Zumino \cite{WZ1,WZ2}.
 
As noted by Misner and Watt \cite{MW}, though scalar gravitational theories are not 
realistic, they have value as computational tools in numerical relativity.  This raises 
a very intriguing question.  The quote, ``History doesn't repeat itself, but it often 
rhymes,\footnote{Though often attributed to Mark Twain, there is little evidence of 
this being accurate.}'' has been stated about many situations.  Since scalar gravitation
models have value as computational tools for General Relativity, might extending
them to eleven and ten dimensional supergeometry offer new ways to replicate 
Einstein's path from the ``happiest thought'' to the higher level of understanding as
indicated by his lectures at G\" ottingen?

It is the purpose of this work to lay a new foundation for such an exploration.  We
take as a guiding principle the procedure used to provide the first example of a
four dimensional supergravity supermultiplet where the closure of the local simple 
supersymmetry algebra was not predicated on the use of field equations.  This
was accomplished by Breitenlohner \cite{B1} who took as his starting point an
off-shell supermultiplet, the so-called ``non-abelian vector supermultiplet.''  From
the vantage of the current time, this is strongly reminiscent of an invocation of
the concept of gravity as the double of gauge theory.  The final form of Breitenlohner
initial presentation realized a reducible representation of supersymmetry.  Finally,
this first work also did not consider any issues surrounding the construction of
actions for the supermultiplet.

With the Breitenlohner approach as a guiding principle to the study of a class of 
curved supermanifolds containing eleven and ten dimensions, it is thus to be 
expected the extensions will manifest the same structure of being off-shell,
but reducible and not address the issue of the construction of actions.  No off-shell
gauge vector supermultiplet is known beyond six dimensions, thus one is forced 
to deviate from completely following the Breitenlohner approach.  Since a scalar 
superfield in any dimension is guaranteed to be off-shell, but reducible, one is 
naturally led to the study of Nordstr\" om supergravity theories in eleven and ten 
dimensions in this approach.

We organize this current paper in the manner described below.

Chapter two provides a self-consistent introduction to the field-theory and
gauge-theory based formulation of gravitation described solely by a
metric in $D$ dimensions.  We use a frame field/spin-connection formulation
from the beginning point of our discussion.  This eases the transition to the
case of superspace for these latter theories as it is an impossibility \cite{W00} to 
introduce a metric/Christoffel Riemannian formulation in the context of a superspace
geometry appropriate to supersymmetry.  The restriction of the full frame field to
retain only the degree of freedom associated with its determinant is presented
along with:
\newline \noindent
$~~~~~$ $~~~~~$  (a.)
the well-known vanishing of the Weyl tensor, and
\newline \noindent
$~~~~~$ $~~~~~$ (b.)
the residual form of the Einstein-Hilbert action under this 
\newline \noindent $~~~~~~\,$ $~~~~~$ $~~~~$ 
restriction.

Chapter three is a transitional one where we review 4D, $\cal N$ = 1
supergravity as a paradigm setting arena.  We show how the structure of
this superspace of this well studied theory suggests pathways that can be
pursued for how to carry out construction of scalar supergravitation in all
higher dimensions including ten and eleven dimensional theories.

In chapters four through seven, we deploy the lessons found in the third 
chapter to work in making respective proposals for linearized theories
of scalar supergravitation in the 11D, $\cal N$ = 1, 10D, $\cal N$ = 2A, 
10D, $\cal N$ = 2B, and 10D, $\cal N$ = 1 superspaces.

We follow this work with our conclusions, two appendices, and the
bibliography.

\newpage
\section{Gauge Theory Perspective On Ordinary Gravity}

The traditional geometrical approach to describing gravity can be regarded as 
having driven an apparent wedge between general relativity and theory of 
elementary particles. Instead, a gauge theory and field theory based point of 
view provides a logical foundation for gravity which permits an alternative to 
geometry. 

For gravitational theories in $D$ dimensions, the gauge group can be taken as the
Poincar\' e group, and the Lie algebra generators are momentum $P_{\un{m}} = 
- i \pa_{\un{m}}$ and spin angular momentum generator $\mathcal{M}_{\un{a}\un{b}}$.
These are taken to satisfy the following commutation relations,
\begin{gather}
[~ P_{\un{m}} \,, \, P_{\un{n}} ~] ~=~ 0  
 \qquad , \qquad  
[~ \mathcal{M}_{\un{a}\un{b}} \, ,  \, P_{\un{m}} ~] ~=~ 0  
 \qquad , \qquad  
[~ \mathcal{M}_{\un{a}\un{b}} \, , \,  \pa_{\un{c}} ~] ~=~ \eta_{\un{c}\un{a}} \pa_{
\un{b}} ~-~ \eta_{\un{c}\un{b}} \pa_{\un{a}} ~~~,  \\
[~ \mathcal{M}_{\un{a}\un{b}} \,  , \,  \mathcal{M}_{\un{c}\un{d}} ~] ~=~ \eta_{\un{
c}\un{a}} \mathcal{M}_{\un{b}\un{d}} ~-~ \eta_{\un{c}\un{b}} \mathcal{M}_{\un{a}\un{
d}} ~-~ \eta_{\un{d}\un{a}} \mathcal{M}_{\un{b}\un{c}} ~+~ \eta_{\un{d}\un{b}} \mathcal{
M}_{\un{a}\un{c}} ~~~,
\end{gather}
and it might appear that the definition $P_{\un{m}} = - i \pa_{\un{m}}$ together with
the second and third equations among (2.1) are in contradiction.  The resolution 
to this conundrum is to note
\be
\pa_{\un{a}} ~\equiv~ \d{}_{\un{a}}{}^{\un{m}} \, \pa_{\un{m}} ~~~,
\ee
and the factor of $\d{}_{\un{a}}{}^{\un{m}} $ actually corresponds to the vacuum 
value of the inverse frame field $e_{\un{a}}{}^{\un{m}}$ whose first index
transforms under the action of $\mathcal{M}_{\un{a}\un{b}}$ and whose second 
index is inert under the action of the  spin angular momentum generator.  To 
distinguish between these two types of quantities, we use the ``early'' latin letters, 
${\un{a}}$, ${\un{b}}$, etc. to denote indices that transform under the action of 
$\mathcal{M}_{\un{a}\un{b}}$.  Similarly, we use the ``late'' latin letters, ${\un{m}}$, 
${\un{n}}$, etc. to denote indices that do {\em {not}} transform under the action of  
$\mathcal{M}_{\un{a}\un{b}}$.

The covariant derivative with respect to this gauge group is
\begin{equation}
\nabla_{\un{a}} ~\equiv~ e_{\un{a}}{}^{\un{m}} \pa_{\un{m}} ~+~ \tfrac{1}{2} 
\omega_{\un{a}\un{c}}{}^{\un{d}} \mathcal{M}_{\un{d}}{}^{\un{c}}  ~~~,
\end{equation}
where $e_{\un{a}}{}^{\un{m}}$ is related to the metric through its inverse $e_{\un{m}}
{}^{\un{a}}$ via $g_{\un{m}\un{n}} = e_{\un{m}}{}^{\un{a}} \eta_{\un{a}\un{b}} e_{\un
{n}}{}^{\un{b}}$. The commutator of $\nabla_{\un{a}}$ generates field strengths 
torsions $T_{\un{a}\un{b}\un{c}}$ and curvatures $R_{\un{a}\un{b}\un{c}\un{d}}$
\begin{equation}
[\,  \nabla_{\un{a}} ~,~ \nabla_{\un{b}} \, ] ~=~ T_{\un{a}\un{b}}{}^{\un{c}} \,
\nabla_{\un{c}} ~+~ \tfrac{1}{2} R_{\un{a}\un{b}\un{c}}{}^{\un{d}}  \,
\mathcal{M}_{\un{d}}{}^{\un{c}}
~~~.
\end{equation}
Scalar gravitation can be defined by restricting the form of the frame field to
\begin{equation}
e_{\un{a}}{}^{\un{m}} ~=~ \psi\,  \delta_{\un{a}}{}^{\un{m}} ~~~,
\end{equation}
where $\psi$ is a finite scalar field.  By definition, this defines a class of geometries
that is conformally flat in the context of strictly Riemannian spaces.  To see this we
begin by setting $T_{\un{a}\un{b}\un{c}} = 0$, which implies
\be
\nabla_{\un a} ~=~\psi \, [ ~ \pa_{\un a} ~-~ \, (\pa_{\un b} \, \ln \psi) {\cal M
}_{\un a} {}^{\un b} ~] ~~~.
\label{ClS} \ee
and allows the full Riemann curvature tensor to be solely 
expressed in terms of the $\psi$ field as
\begin{equation}
R_{\un{a}\un{b}}^{\ \ \un{c}\un{d}} ~=~ -\psi(\pa_{[\un{a}}\partial^{[\un{c}}\psi)
\delta_{\un{b}]}^{\ \un{d}]} ~+~ (\partial^{\un{e}}\psi)(\pa_{\un{e}}\psi)
\delta_{[\un{a}}^{\ \un{c}}\delta_{\un{b}]}^{\ \un{d}}  ~~~, 
\end{equation}
similarly for the Ricci curvature we find
\begin{equation}
R_{\un{a}}^{\ \un{c}} ~=~ R_{\un{a}\un{b}}^{\ \ \un{c}\un{b}} ~=~ -(D-2)\psi(\pa_{
\un{a}}\partial^{\un{c}}\psi)-\psi(\Box \psi)\delta_{\un{a}}^{\ \un{c}} ~+~ (D-1)(
\partial^{\un{e}}\psi)(\pa_{\un{e}}\psi)\delta_{\un{a}}^{\ \un{c}} ~~~,
\end{equation}
and finally for the curvature scalar by $R = \delta_{\un{c}}^{\ \un{a}}R_{\un{a}}^{\ \un{c}}$,
\begin{equation}
R ~=~ - 2 (D-1) \psi  (\Box\psi) ~+~ D (D-1) (\partial^{\un{e}}\psi) (\pa_{\un{e}}\psi)
~~~.
\end{equation}
The Weyl tensor ${\cal C}_{\un a \, \un b}{}^{\un c \, \un d}$ is defined by the equation
\be 
{\cal C}_{\un a \, \un b}{}^{\un c \, \un d} 
~=~ R_{\un a \, \un b}{}^{\un c \, \un d} ~-~ \Big[ {1 \over {D - 2}}\Big] 
R_{[ {\un a}}{}^{[{\un c }} \d{}_{{\un b}]}{}^{ { \un d} ]}  ~+~ 
\Big[{1 \over {(D - 2})({D - 1})} \Big] \, \delta_{[\un{a}}^{\ \un{c}}
\delta_{\un{b}]}^{\ \un{d}} \, R
~~~,
\label{WyL}
\ee
and when the results in (2.8) - (2.10) are used, this is found to vanish.

We define $e \equiv \det (e_{\un{a}}{}^{\un{m}}) = \psi^{D}$ and the Einstein-Hilbert action 
takes the form
\begin{equation}
\eqalign{  {~~~~}
S_{EH} ~&=~  \frac{3}{\kappa^{2}} \int d^{D}x \; e^{-1} R(\psi) =  \frac{3}{\kappa^{2}} 
\int d^{D}x \, \psi{}^{-D} \, \; \big[ - 2 (D-1) \psi  (\Box\psi) + D (D-1) (\partial^{\un{e}}\psi)
(\pa_{\un{e}}\psi) \big]   \cr
~&=~   \frac{3}{\kappa^{2}} \int d^{D}x \,\; \big[ - 2 (D-1) \,  \psi{}^{1-D}  \,  (\Box\psi) + 
D (D-1)\,  \psi{}^{-D} \,  (\partial^{\un{e}}\psi) (\pa_{\un{e}}\psi) \big]     \cr
~&=~ - \,   \frac{3}{\kappa^{2}} \, (D-1)  \int d^{D}x \, \big\{ \, ( D- 2 ) \, \big[ \, \psi{}^{-D} 
\, (\partial^{\un{e}}\psi) (\pa_{\un{e}}\psi) \big] ~+~ 2 \, \pa^{\un e} \big[ \, \psi{}^{1-D}
\, (\pa_{\un e}\psi) \big] \, \big\}  ~~~.
} \end{equation}

As the full off-shell description of 10D and 11D supergravities are yet unknown, we work with 
a toy model - scalar supergravity in the higher dimensions, which we expect gives part of the 
complete solutions. In the subsequent chapters, we replace $\psi$ by $1+\Psi$, where $\Psi$ is an infinitesimal 
superfield, and study the corresponding linearized supergravity.

\newpage
\section{Nordstr\" om Supergravity in 4D, $\mathcal{N}=1$ Supergeometry}

As a preparatory step for our eventual goals, it is important that we re-visit four dimensional
$\cal N$ = 1 linearized supergravity as there are important lessons to be gained from
asking questions solely in this domain prior to making the leap to eleven and ten dimensions.  
The formulation of linearized 4D, $\cal N$ = 1 supergravity in term of the usual supergravity 
pre-potential $H^{\un{a}}$ was identified long ago \cite{OS}.  It is perhaps of importance 
to note that supergravity pre-potentials bare some resemblance to other better known 
concepts important for the mathematical description of theories describing gravitation.

One of the most computationally enabling formulations of the dynamics of ordinary gravitation
is based on the Arnowitt-Deser-Meisner (ADM) formulation \cite{ADM} wherein the
quadratic form involving the metric takes the form
\be  \eqalign{  {~~~}
d x^{\un m} \, g_{\un m  \, \un n}  \, d x^{\un n}  ~&=~
\left( \, - N{}^2 ~+~ N{}_{\un i} \d{}^{{\un i} \, {\un j}}  N{}_{\un j} \, \right) dt \otimes 
d t ~+~  N{}_{\un i} \, \left( \,  dt \otimes d x{}^{{\un i}}  \,+\,  dx{}^{{\un i}} \otimes dt \, 
\right) ~+~ {\Sc g} {}_{{\un i} \, {\un j}} \, d x{}^{{\un i}} d x{}^{{\un j}}   ~~~,
}
\label{eq:1Chp1} 
\ee
in terms of the ``lapse'' function $N$, ``shift'' vector $N{}_{\un i} $, and induced
3-metric ${\Sc g} {}_{{\un i} \, {\un j}}$.  For the equation above to be valid, we can write
\begin{equation}
\begin{aligned}
 d x^{\un m} ~=~ \left(  \, c\, dt , ~ dx{}^{\un 1} , ~ dx{}^{\un 2} , ~ dx{}^{\un 3}   \, \right) ~~,~~
g{}_{{\un m} \, {\un n}}  
~=&~ {\begin{bmatrix}
\fracm 1{c^2} \, [ \,  - N{}^2 ~+~ N{}_{\un i} \d{}^{{\un i} \, {\un j}}  N{}_{\un j}  \, ] & 
~~  \fracm 1{c} \, N{}_{\un 1}~~ &  ~~ \fracm 1{c} \, N{}_{\un 2} ~~&  ~~ \fracm 1{c} \, N{}_{
\un 3} ~~ \\
{} & {} & {} & {} \\
\fracm 1{c} \, N{}_{\un 1} & {\Sc g} {}_{ {\un 1} \,  {\un1}} & {\Sc g} {}_{ {\un1} \,  {\un 2}} 
& {\Sc g} {}_{ {\un1}  {\un 3}} \\
{} & {} & {} & {} \\   
\fracm 1{c} \, N{}_{\un 2} & {\Sc g} {}_{ {\un 2} \,  {\un 1}} & {\Sc g} {}_{ {\un 2} \,  {\un 2}} 
& {\Sc g} {}_{ {\un 2} \,  {\un 3}} \\ {} & {} & {} & {} \\  
\fracm 1{c} \, N{}_{\un 3} & {\Sc g} {}_{{\un 3} \, {\un 1}} & {\Sc g} {}_{{\un 3} \, {\un 2}} & 
{\Sc g} {}_{{\un 3} \, {\un 3}} \\
\end{bmatrix}} 
~~.
\end{aligned}
\label{eq:2Chp2}
\end{equation}
The introduction of frame fields can be accomplished by observing that the quadratic form in 
(\ref{eq:1Chp1}) may also be written as
\be 
d x^{\un m} \, g_{\un m  \, \un n}  \, d x^{\un n}  ~\equiv~ d x^{\un m} \, e_{\un m}{}^{\un a} 
\, \eta_{\un a \un b} \, e_{\un n} {}^{\un b} \, d x^{\un n}  {~~~},
\label{eq:2Chp15} \ee
by ``factorizing'' the metric into the product of two frame fields $e_{\un m}{}^{\un a}(x)$ 
and $e_{\un n}{}^{\un b}(x)$ multiplied by the constant Minkowski metric, $\eta_{\un a 
\un b}$, of flat spacetime.  Thus, there exist relations between the ADM variables and
the frame fields  \cite{ADMv,ADMv2}.

The point of the above discussion is to note that the inverse frame fields $e{}_{\un a
}{}^{ \un m} $ may be regarded as functions of the ADM variables, i.\ e.\
\be
e{}_{\un a}{}^{ \un m} ~=~ e{}_{\un a}{}^{ \un m}\big( N, \, N{}_{\un i}, \,
{\Sc g} {}_{{\un i} \, {\un j}} \big) ~~~,
\ee
and that for numerical relativity calculations, the latter are far more useful than the
frame fields $e{}_{\un m}{}^{ \un a}$, or even the metric $g{}_{{\un m} \, {\un n}}$ itself.
As we will see later, it is the form of 4D, $\cal N$ = 1 supergravity often called the
``Breitenlohner auxiliary field set'' that is relevant to our work.  For this formulation, 
it was first shown in the work of \cite{SFSG} the super-frame superfields $E{}_{\un 
A}{}^{ \un M}$ are expressed in terms of a more fundamental set of superfields, i.e. 
the prepotentials $H^{\un{m}}$ and $\Psi$ (with the ``conformal compensator'' explicitly
dependent upon a complex linear superfield).  In an 
``echo'' of the utility of the ADM variables, the prepotentials are far more useful 
when component calculations, or quantum calculations are undertaken, with the 
latter able to utilize the technology of super Feynman graphs.

As in the discussion of section 7.5 in \cite{SSp8cBK}, we write (with $X$ being the 
superfield linearization of
$\Psi$)
\begin{equation}
\begin{split}
{\rm E}_{\alpha} ~=&~  {\rm D}_{\alpha} + \Bar{X} {\rm D}_{\alpha} + i \frac 12 \,
({\rm D}_{\alpha} H^{\un{b}}) \pa_{\un{b}} ~~~,   \\
{\rm E}_{\Dot{\alpha}} ~=&~  \Bar{{\rm D}}_{\Dot{\alpha}} + X \Bar{{\rm D}}_{\Dot{\alpha}}
- i \frac 12 \, ( \Bar{{\rm D}}_{\Dot{\alpha}} H^{\un{b}}) \pa_{\un{b}} ~~~,
\end{split}
\end{equation}
\begin{equation}
\begin{split}
{\rm E}_{\un{a}} ~=&~ \pa_{\un{a}} + i \Big[ \frac{1}{2} \Bar{{\rm D}}^{2} {\rm D}_{(\alpha} 
H^{\gamma)}{}_{\Dot{\alpha}} - (\Bar{{\rm D}}_{\Dot{\alpha}} \Bar{X}) \delta_{\alpha}{}^{
\gamma} \Big] {\rm D}_{\gamma} + i \Big[ - \frac{1}{2} {\rm D}^{2} \Bar{{\rm D}}_{(\Dot{
\alpha}} H_{\alpha}{}^{\Dot{\gamma})} - ({\rm D}_{\alpha} X)  \delta_{\Dot{\alpha}}{}^{
\Dot{\gamma}} \Big] \Bar{{\rm D}}_{\Dot{\gamma}}  \\
&  + \Big[ -\, \frac 12 ( \, [ {\rm D}_{\alpha} ~,~ \Bar{{\rm D}}_{\Dot{\alpha}} ] H^{\un{b}}) + 
(X + \Bar{X}) \delta_{\un{a}}{}^{\un{b}} \Big] \pa_{\un{b}} ~~~.
\end{split}
\end{equation}
for the linearized superframe superfield.   Similar to the ADM formulation of
ordinary gravity, the superframe is expressed in terms of two independent
superfields, $H^{\un{a}}$, and $X$.  The remaining structures needed to
specify the supergravity supercovariant derivative are the spin-connections
which here take the forms
\begin{equation}
\begin{split}
\Phi_{\alpha\beta\gamma} ~=&~ - C_{\alpha(\beta} {\rm D}_{\gamma)} \Bar{X}  ~~~, \\
\Phi_{\alpha\Dot{\beta}\Dot{\gamma}} ~=&~ \frac{1}{2} {\rm D}^{2} \Bar{{\rm D}}_{(\Dot{\beta}|} H_{\alpha|\Dot{\gamma})}  ~~~, \\
\Phi_{\Dot{\alpha}\beta\gamma} ~=&~ - \frac{1}{2} \Bar{{\rm D}}^{2} {\rm D}_{(\beta} 
H_{\gamma)\Dot{\alpha}}  ~~~, \\
\Phi_{\un{a}\beta\gamma} ~=&~ i \frac{1}{2} {\rm D}_{\alpha} \Bar{{\rm D}}^{2} {\rm D}_{
(\beta} H_{\gamma)\Dot{\alpha}} + i C_{\alpha(\beta|} \Bar{{\rm D}}_{\Dot{\alpha}} {\rm 
D}_{|\gamma)} \Bar{X} ~~~.
\end{split}
\end{equation}

The superfield $X$ introduced above is a general scalar superfield.  This implies
that the linearized formulation described above {\em {is}} reducible.  There 
are two widely familiar choices that lead to irreducibility.  One choice is implemented
by picking $X$ to depend on $H^{\un{a}}$, and a chiral superfield $\phi$ (i.\ e.\
${\Bar{\rm D}}{}_{\Dot \a} \phi = 0$).  This is the path that leads to the minimal off-shell
formulation of 4D, $\cal N$ = 1 supergravity.  For this choice, the commutator 
algebra of the superspace supergravity covariant derivative takes the forms
\be
\eqalign{ 
[ \nabla_{\a }, \nabla_{\b} \}  &=~ - 2 \Bar R \,{ \cal M}_{\a \b}
~~~~~, ~~~~~ 
[ \nabla_{\a }, { \Bar {\nabla}}_{\Dot \a} \}  ~=~ i {\nabla}_{ \un a}
~~~~~, \cr
[ \nabla_{\a }, \nabla_{\un b} \}  &=~ - i \, C_{ \a \b} \Big[~ \Bar R \, 
{\Bar {\nabla}}_{\Dot \b} ~-~ G^{\g} {}_{\Dot \b } \nabla_{\g} ~ \Big] 
~-~ i \, ( {\Bar \nabla}_{\Dot \b} \Bar R )  { \cal M}_{\a \b} ~~ \cr
&~~~~~+~ i \, C_{\a \b} \Big[~ {\Bar W}_{\Dot \b \Dot \g} {}^{\Dot \d}
 { \Bar {\cal M}}_{\Dot
\d} {}^{\Dot \g} ~-~ ( \nabla^{\d} G_{\g \Dot \b}) {\cal M}_{
\d} {}^{ \g}  ~ \Big]  ~~~, \cr
[ {\nabla}_{\un a }, \nabla_{\un b} \}  &=~ \Big\{ ~ \Big[~ C_{\Dot \a 
\Dot \b} W_{ \a \b} {}^{ \g} 
~+~ \fracm 12  C_{ \a \b} ({\Bar \nabla}_{(\Dot \a} G^{\g}{}_{ \Dot \b )} ) ~-~ \fracm 12 C_{\Dot \a \Dot \b} \,( 
\nabla_{(\a} R  \,)\d_{\b)}^{\ \g}  \,~ \Big] \nabla_{\g} \cr
&~~~~~+~  i \fracm12  C_{\a \b } G^{\g} {}_{ ( \Dot \a|} \nabla_{ \g |\Dot \b)} \cr
&~~~~~-~ \Big[~ C_{\Dot \a \Dot \b} \Big( \nabla_{\alpha} W_{\b \d \g} ~+~  \fracm 12  C_{\d ( \a } C_{\b ) \g } \, 
( \, {\Bar \nabla}^2 \Bar R ~+~ 2 R \Bar R) ~ \Big) \cr
&~~~~~-~ \fracm 12 C_{\a \b} \, ({\Bar \nabla}_{( \Dot \a|} \nabla_{\g} G_{\d | \Dot \b )}) ~\Big] 
{\cal M}^{\g \d}  ~ \Big\} ~+~ 
{\rm{h.\, c.}}   ~~~. }\label{eq:uzw}
\ee

The other widely known choice ``the Breitenlohner auxiliary field set'' is 
implemented by picking $X$ to depend on $H^{\un{a}}$, and a complex linear 
superfield $\Sigma$ (i.\ e.\ ${\Bar {\rm D}}{}^2 \Sigma = 0$).  This is the path 
that leads to the non-minimal off-shell formulation of 4D, $\cal N$ = 1 supergravity.  
For this choice, the commutator algebra of the superspace supergravity 
covariant derivative takes the forms
\begin{equation}
\eqalign{  {~~}
[ \de_{\a} ~,~ \de_{\b} \} ~=~ &\frac 12 T_{(\a} \de_{\b)} ~-~
2 \,{\Bar R} {\cal M}_{\a \b}  ~~~, \cr
[ \de_{\a} ~,~ {\Bar \de}_{\Dot \b} \} ~=~ &i  \, \de_{\a \Dot \b}  ~~~, \cr
[ \de_{\a} ~,~ \de_{\un b} \} ~=~ &\frac 12 \,T_{\b} \de_{\a \Dot \b}  ~-~iC_{\a\b} \Big[ \Bar{R} ~+~ \frac{1}{4} ( \nabla^{\g}T_{\g} ) \Big] {\Bar \nabla}_{\Dot \b} \cr
&+~ i \,\Big[~ C_{\a \b} \, G^{\g} {}_{\Dot \b} ~-~ \frac{1}{2}C_{\a\b} \big(  (\nabla^{\g} ~+~ \fracm{1}{2}T^{\g})\Bar{T}_{\Dot\b}  \big)
~+~ \frac 12 (\, {\Bar \de}_{\Dot \b} {T}_{\b}\, ) \d_{\a}{}^{\g}  ~\Big] \de_{\g} \cr
&-~ i \, \Big[~ C_{\a \b} \, (\de^{\g} G_{\d \Dot \b} ) {\cal M}_{\g}{}^{\d}
~+~  \big(\, ({\Bar \de}_{\Dot \b}-\Bar{T}_{\Dot\b} ) {\Bar R} \,\big) {\cal M}_{\a \b} ~\Big] \cr
&+~ i \, C_{\a \b} \, \Big[~  {\Bar W}_{\Dot \b \Dot \g}{}^{\,\Dot \d} \,
{\Bar {\cal M}}_{\Dot \d }{}^{\Dot \g} ~+~  \frac 16 \, \big( \de^{\d}\,  \,( \de_{\d} ~+~ \fracm 12 T_{\d} \, ) \,   {\Bar  T}_{\Dot \g} \,\big) ~
{\Bar {\cal M}}_{\Dot \b}{}^{\Dot \g}  ~+~  \frac 13 \, {\Bar R} \, {\Bar  T}_{\Dot \g} \, 
{\Bar {\cal M}}_{\Dot \b}{}^{\Dot \g} ~\Big] ~~~. }
\end{equation}
The final commutator $[ \de_{\un a} ~,~ \de_{\un b} \}$ is found to be
explicitly found from the equation 
\begin{equation}
\eqalign{  {~~}
[ \de_{\un a} ~,~ \de_{\un b} \} 
=~ &  \Big\{\, i \frac 12 \, ( \de_{\b} {\Bar T}_{\Dot\b} ) \de_{\un a} ~-~ i \frac 12 \, ( {\Bar \de}_{\Dot \a} T_{\b} ) \de_{\a \Dot \b} ~-~ i \, C_{\Dot\a \Dot\b} \, \Big[~  G_{\b} {}^{\Dot \g} ~+~ \frac12 \, \big( ~ ( {\Bar \de}^{\Dot \g} ~+~ \fracm12 {\Bar T}^{\Dot \g} ) \, T_{\b} ~ \big) ~\Big] \de_{\a \Dot \g}  \cr 
& +~   \frac 12 \, \Big[~ ( {\Bar \de}_{\Dot \a} {\Bar \de}_{\Dot \b} T_{\b} ) ~-~ \frac 12 \, {\Bar T}_{\Dot \b} \, ( {\Bar \de}_{\Dot \a} T_{\b} ) ~\Big] ~ \de_{\a}     \cr
& -~  C_{\Dot\a \Dot\b} \, \Big[~ ( \de_{\a} R ) ~+~ \frac14 \, (\de_{\a} {\Bar\de}^{\Dot\g} {\Bar T}_{\Dot\g} ) ~+~ \frac13 \, R \, T_{\a} ~-~ \frac{1}{12} \, \big( ~ {\Bar\de}^{\Dot\g} \, ( {\Bar\de}_{\Dot\g} ~-~ {\Bar T}_{\Dot\g} ) \, T_{\a} ~ \big) ~ \Big] \, \de_{\b}     \cr
& +~  C_{\Dot\a \Dot\b} \,  W_{\b \a}{}^{\, \g} \, \de_{\g} ~+~  C_{\a\b} \, \Big[~ ( {\Bar \de}_{\Dot \a}  G^{\g} {}_{\Dot \b} ) ~-~ \frac 12 \, {\Bar T}_{\Dot \b} \,  G^{\g} {}_{\Dot \a}  ~\Big] \,  \de_{\g}   ~+~  \frac12 \, C_{\a \b} \,  \Big[~ \frac12 \, ( \de^{\g} {\Bar T}_{\Dot \a} ) \, {\Bar T}_{\Dot \b} \cr 
& \qquad  -~ \big(~ {\Bar \de}_{\Dot \a} \, (\de^{\g} ~+~ \fracm12 T^{\g} ) \, {\Bar T}_{\Dot\b} ~\big) ~+~ \frac 12 \, \big(~ ( {\Bar \de}_{\Dot \a} ~+~ \fracm12 {\Bar T}_{\Dot\a} ) \, {\Bar T}_{\Dot\b} ~\big) \, T^{\g} ~\Big] \,  \de_{\g}  \cr 
& +~ \Big[~ -~ ( {\Bar\de}_{\Dot \a} {\Bar\de}_{\Dot \b} {\Bar R} ) ~+~ \frac 12  ( {\Bar\de}_{\Dot \a} {\Bar R} ) \, {\Bar T}_{\Dot \b} ~+~  {\Bar R} \, ( {\Bar \de}_{\Dot\a} + \fracm12 {\Bar T}_{\Dot\a} ) \, {\Bar T}_{\Dot \b}  ~ \Big] \, {\cal M}_{\a\b}  \cr
& \qquad  +~ 2 \, C_{\Dot\a\Dot\b} \, {\Bar R} \, \Big[~ R ~+~ \frac 14  ( {\Bar\de}^{\Dot \g} {\Bar T}_{\Dot \g} )  ~ \Big] \, {\cal M}_{\a\b}  \cr
& -~ \frac 16 \,  C_{\Dot\a \Dot\b} \, \Big[~  R \, T_{\b} \, T_{\g} ~+~ \frac12 \, T_{\b} \, ( {\Bar\de}^{\Dot\d} {\Bar\de}_{\Dot\d} T_{\g} ) ~+~ \frac14 \, T_{\b} \, T_{\g} \, ( {\Bar\de}^{\Dot\d} {\Bar T}_{\Dot\d} )   ~\Big]  ~ {\cal M}_{\a}{}^{\g}  \cr 
& +~ \frac 16 \,  C_{\Dot\a \Dot\b} \, \Big[~ 2 \, (\de_{\a} R) \, T_{\g} ~+~ 2 \, R \, (\de_{\a} T_{\g} ) ~+~  \big(~ \de_{\a} {\Bar\de}^{\Dot \d} \, ( {\Bar\de}_{\Dot \d} ~+~ \fracm12 {\Bar T}_{\Dot \d} ) \, T_{\g} ~\big)   \cr
& \qquad +~ \fracm 12 ( {\Bar \de}^{\Dot\d} {\Bar T}_{\Dot \d} ) \, ( \de_{\a} T_{\g} )   ~\Big]  ~ {\cal M}_{\b}{}^{\g}  \cr 
& -~  C_{\a\b} \, \Big[~ ( {\Bar \de}_{\Dot \a} \de^{\g} 
G_{\d \Dot \b}  )  ~-~ \frac 12  \, {\Bar T}_{\Dot \b} \, ( \de^{\g} G_{\d \Dot \a}  )  ~\Big] \,  {\cal M}_{\g}{}^{\d}  \cr
& \qquad +~ C_{\Dot\a \Dot\b} \, \Big[~  ( \de_{\a} W_{\b\d}{}^{\, \g} ) ~-~ \frac 12 \, T_{\b} \, W_{\a\d}{}^{\, \g} ~\Big] \,  {\cal M}_{\g}{}^{\d}  \Big\}   \cr 
& +~ \text{h. c.}
~~~. }
\end{equation}

Under either choice, one can use the definitions of the superframe superfields
in (3.5) - (3.7) together with the set of equations of {\em {either}} (3.8) or (3.9) and (3.10) to
find the dependence of $W_{ \a \b \g} $, $G{}_{\un a} $, and ${\Bar R}$ (for the 
minimal theory) on $H{}^{\un a} $, and $\phi$, or  the dependence of $W_{\a \b 
\g}$, $G{}_{\un a} $, ${\Bar R}$, and $T{}_{\a}$ on  $H{}^{\un a} $, and $\S$ (for 
the non-minimal theory).  These are the standard and well discussed theories
of off-shell 4D, $\cal N$ = 1 supergravity, i.\ e.\ the consistency of the Bianchi 
identities associated (3.8) or (3.9) and (3.10) for the algebra of the superspace supergravity 
covariant derivatives do {\em {not}} require on-shell conditions to be imposed 
on the components fields contained within the superfields.

The process of imposing the Einstein Field Equations in the non-supersymmetrical
case in the absence of matter amounts to the condition
\be
{R}_{\un a \, \un b} ~-~ \fracm 12 \eta_{\un a \un b}
\, {R} ~+~ \Lambda \eta_{\un a \un b} ~=~  0  ~~~,
\ee
including the cosmological constant.   The equivalent in the case of superspace 
supergravity is accomplished by setting  $G{}_{\un a} $, and ${\Bar R}$ (for the 
minimal theory) to zero or by setting $G{}_{\un a} $, and $T{}_{\a}$  (for the 
non-minimal theory) to zero.  The condition $T{}_{\a}$ = 0 also forces ${\Bar R}$ 
= 0 in the non-minimal theory.  Under these conditions, the algebra of superspace
supergravity covariant derivatives take the universal form
\begin{equation}
\eqalign{  {~~}
[ \nabla_{\a }, \nabla_{\b} \}  &=~ 0 ~~~~~, ~~~~~ 
[ \nabla_{\a }, { \Bar {\nabla}}_{\Dot \a} \}  ~=~ i {\nabla}_{ \un a}
~~~~~, \cr
[ \nabla_{\a }, \nabla_{\un b} \}  &=~  i \, C_{\a \b} [~ {\Bar W}_{\Dot \b \Dot \g} 
{}^{\Dot \d} { \Bar {\cal M}}_{\Dot
\d} {}^{\Dot \g}  ~ ]  ~~~, \cr
[ \de_{\underline a} ~,~ \de_{\un b} \} 
&=~   C_{\Dot\a \Dot\b} \,  W_{\b \a}{}^{\, \g} \,    \de_{\g}   +~    C_{\Dot\a \Dot\b} \, (
 \de_{\a} W_{\b\g}{}^{\, \d} ) \, {\cal M}_{\d}{
}^{\g}   ~+~ \text{h. c.}
~~~~. }
\end{equation}

At this point, we can take a largely unexplored path as it is possible to consider 
the limit of these equations wherein $H{}^{\un a} $ = 0.  This is the route to the 
4D, $\cal N$ = 1 superspace version of scalar supergravitation theory \`a la 
Nordstr\" om in the eleven and ten dimension that are the targets of our study.

The curious reader may wonder from where does the condition $H{}^{\un a} $ 
= 0 arise?  On page 473, of \cite{SSp8cBK} there appears the following text.

$~~~~$ {\it {Nonsupersymmetric deSitter covariant 
derivatives can be obtained from gravitational}} 
\newline \indent 
{\it {covariant 
derivatives by eliminating all field components except the 
(density) compensat}} \newline \indent 
{\it {-ing field (i.e., the determinant of the 
metric or vierbein). This follows from the fact that}}
\newline \indent 
{\it {in deSitter
space the Weyl tensor vanishes, which says that there is no 
conformal (spin 2) }}
\newline \indent 
{\it {part to the metric: It is ÔÔconformally flatÕÕ.}} 

\noindent From the discussion given in chapter two, we saw that Nordstr\" om geometries
in all dimensions are necessary such as to describe Weyl tensors that vanish
and are hence conformally flat.

Also in the work of \cite{SSp8cBK} it is explained that the conformal part of the 
metric arises solely from $H{}^{\un a} $. Since the Nordstr\" om limit is a conformally 
flat bosonic space, it must corresponds to setting $H{}^{\un a} $ = 0. Thus, to our 
knowledge the passage above from ``Superspace'' marked the first indication of 
this.  Of course, other authors such as in the work of \cite{B&K} later reaffirmed this
point about the structure of superspace of supergravity.

In the limit of our interest, we find
\begin{equation}
\begin{split}
{\rm E}_{\alpha} ~=&~  {\rm D}_{\alpha} + \Bar{X} {\rm D}_{\alpha} ~~~,~~~
{\rm E}_{\Dot{\alpha}} ~=~  \Bar{{\rm D}}_{\Dot{\alpha}} + X \Bar{{\rm D}}_{\Dot{\alpha}}
~~~, \\
{\rm E}_{\un{a}} ~=&~ \pa_{\un{a}} - i \Big[  
(\Bar{{\rm D}}_{\Dot{\alpha}} \Bar{X}) \delta_{\alpha}{}^{
\gamma} \Big] {\rm D}_{\gamma} - i \Big[  ({\rm D}_{\alpha} X)  \delta_{\Dot{\alpha}}{}^{
\Dot{\gamma}} \Big] \Bar{{\rm D}}_{\Dot{\gamma}}  
+ \Big[  (X + \Bar{X}) \delta_{\un{a}}{}^{\un{b}} \Big] \pa_{\un{b}} ~~~, \\
\Phi_{\alpha\beta\gamma} ~=&~ - C_{\alpha(\beta} {\rm D}_{\gamma)} \Bar{X}  ~~~,~~~
\Phi_{\alpha\Dot{\beta}\Dot{\gamma}} ~=~ 0  ~~~, ~~~
\Phi_{\Dot{\alpha}\beta\gamma} ~=~ 0  ~~~, ~~~
\Phi_{\un{a}\beta\gamma} ~=~ i C_{\alpha(\beta|} \Bar{{\rm D}}_{\Dot{\alpha}} {\rm 
D}_{|\gamma)} \Bar{X} ~~~.
\end{split}
\end{equation}

In response to this restriction, the forms of the algebras in (3.8), (3.9) and (3.10) also
change.  In particular, the superfield $W_{ \a \b \g} $ (and consequently
$W_{ \a \b \g \d} $) is identically zero.  The latter condition is consistent with
the component level description of scalar gravitation in the previous chapter
as the Weyl tensor of (2.11) is the leading component field that occurs in
$W_{ \a \b \g \d} $ and occurs at first order in the $\theta$-expansion of
$W_{ \a \b \g} $.
The third result in (3.13) also contains two useful bits of information:
\newline \indent
(a.) The
final term of the equation informs us that the leading term in the $\theta$-expansion
\newline \indent
$~~~~\,~$ 
of $X + {\Bar X}$ corresponds to the linearization of $\psi$ seen in equation
(2.6).
\newline \indent
(b.)
The
second term of the equation informs us that the leading term  in the $\theta$-expansion
\newline \indent
$~~~~\,~$ 
of $ (\Bar{{\rm D}}_{\Dot{\alpha}} \Bar{X}) \delta_{\alpha}{}^{
\gamma}$ corresponds to the spin-1/2 remnant of the gravitino!

Another point to discuss is the dependence of the field strength superfields $G{}_{\un a} 
$, and ${\Bar R}$ (for the minimal theory) and  $G{}_{\un a} $, ${\Bar 
R}$, and $T{}_{\a}$ (for the non-minimal theory) on the superfield $X$.  Direct
calculation shows that the reality of $G{}_{\un a}$ in both cases implies that
it only depends on the difference $i(X - {\Bar X})$.  The superfield $T{}_{\a}$ 
is found to depend on the first spinor derivative (i.\ e.\ ${\rm D}{}_{\a}$) of $X$.  
Finally, the superfield ${\Bar R}$ is found to depend on the second spinorial 
derivative of an expression linear in $X$ and $\Bar X$.

We have argued previously \cite{M2}, the minimal supergravity theory does 
not extend from four dimensions to eleven dimensions since there is no concept 
of chirality in the higher dimension.  This implies that only the features seen in 
the non-minimal theory should be expected to occur in the subsequent chapters 
of this work.  As we shall see, this is indeed the case.  The commutator algebra 
for the superspace supergravity covariant derivative responds to the condition 
$H{}^{\un a}$ = 0, by the elimination of the all terms proportional to $W{}_{\a \b 
\g}$ and $W{}_{\a \b \g \d}$.  Thus, we find 4D, $\cal 
N$ = 1 Nordstr\" om supergravity {\em {that}} {\em {descends}} from ten
or eleven dimensions and only contains the generators associated with
4D, $\cal N$ = 1 simple supegravity
is described by
$$
\eqalign{  {~~}
[ \de_{\a} ~,~ \de_{\b} \} ~=&~ \frac 12 T_{(\a} \de_{\b)} ~-~
2 \,{\Bar R} {\cal M}_{\a \b} ~~~, \cr
[ \de_{\a} ~,~ {\Bar \de}_{\Dot \b} \} ~=&~ i  \, \de_{\a \Dot \b}~~~, 
\cr  
[ \de_{\a} ~,~ \de_{\un b} \} ~=&~ \frac 12 \,T_{\b} \de_{\a \Dot \b}  ~-~iC_{\a\b} \Big[ \Bar{R} ~+~ \frac{1}{4} ( \nabla^{\g}T_{\g} ) \Big] {\Bar \nabla}_{\Dot \b} \cr
&+~ i \,\Big[~ C_{\a \b} \, G^{\g} {}_{\Dot \b} ~-~ \frac{1}{2}C_{\a\b} \big(  (\nabla^{\g} ~+~ \fracm{1}{2}T^{\g})\Bar{T}_{\Dot\b}  \big)
~+~ \frac 12 (\, {\Bar \de}_{\Dot \b} {T}_{\b}\, ) \d_{\a}{}^{\g}  ~\Big] \de_{\g} \cr
&-~ i \, \Big[~ C_{\a \b} \, (\de^{\g} G_{\d \Dot \b} ) {\cal M}_{\g}{}^{\d}
~+~  \big(\, ({\Bar \de}_{\Dot \b}-\Bar{T}_{\Dot\b} ) {\Bar R} \,\big) {\cal M}_{\a \b} ~\Big] \cr
&+~ i \, C_{\a \b} \, \Big[~  \frac 16 \, \big( \de^{\d}\,  \,( \de_{\d} ~+~ \fracm 12 T_{\d} \, ) \,   {\Bar  T}_{\Dot \g} \,\big) ~
{\Bar {\cal M}}_{\Dot \b}{}^{\Dot \g}  ~+~  \frac 13 \, {\Bar R} \, {\Bar  T}_{\Dot \g} \, 
{\Bar {\cal M}}_{\Dot \b}{}^{\Dot \g} ~\Big] ~~~,  \cr 
} $$

\begin{equation}
\eqalign{  {~~}
[ \de_{\un a} ~,~ \de_{\un b} \} 
~=&~   \Big\{\,i \frac 12 \, ( \de_{\b} {\Bar T}_{\Dot\b} ) \de_{\un a} ~-~ i \frac 12 \, ( {\Bar \de}_{\Dot \a} T_{\b} ) \de_{\a \Dot \b} ~-~ i \, C_{\Dot\a \Dot\b} \, \Big[~  G_{\b} {}^{\Dot \g} ~+~ \frac12 \, \big( ~ ( {\Bar \de}^{\Dot \g} ~+~ \fracm12 {\Bar T}^{\Dot \g} ) \, T_{\b} ~ \big) ~\Big] \de_{\a \Dot \g}  \cr 
& +~   \frac 12 \, \Big[~ ( {\Bar \de}_{\Dot \a} {\Bar \de}_{\Dot \b} T_{\b} ) ~-~ \frac 12 \, {\Bar T}_{\Dot \b} \, ( {\Bar \de}_{\Dot \a} T_{\b} ) ~\Big] ~ \de_{\a}     \cr
& -~  C_{\Dot\a \Dot\b} \, \Big[~ ( \de_{\a} R ) ~+~ \frac14 \, (\de_{\a} {\Bar\de}^{\Dot\g} {\Bar T}_{\Dot\g} ) ~+~ \frac13 \, R \, T_{\a} ~-~ \frac{1}{12} \, \big( ~ {\Bar\de}^{\Dot\g} \, ( {\Bar\de}_{\Dot\g} ~-~ {\Bar T}_{\Dot\g} ) \, T_{\a} ~ \big) ~ \Big] \, \de_{\b}     \cr
& +~  C_{\a\b} \, \Big[~ ( {\Bar \de}_{\Dot \a}  G^{\g} {}_{\Dot \b} ) ~-~ \frac 12 \, {\Bar T}_{\Dot \b} \,  G^{\g} {}_{\Dot \a}  ~\Big] \,  \de_{\g}   ~+~  \frac12 \, C_{\a \b} \,  \Big[~ \frac12 \, ( \de^{\g} {\Bar T}_{\Dot \a} ) \, {\Bar T}_{\Dot \b} \cr 
& \qquad  -~ \big(~ {\Bar \de}_{\Dot \a} \, (\de^{\g} ~+~ \fracm12 T^{\g} ) \, {\Bar T}_{\Dot\b} ~\big) ~+~ \frac 12 \, \big(~ ( {\Bar \de}_{\Dot \a} ~+~ \fracm12 {\Bar T}_{\Dot\a} ) \, {\Bar T}_{\Dot\b} ~\big) \, T^{\g} ~\Big] \,  \de_{\g}  \cr 
& +~ \Big[~ -~ ( {\Bar\de}_{\Dot \a} {\Bar\de}_{\Dot \b} {\Bar R} ) ~+~ \frac 12  ( {\Bar\de}_{\Dot \a} {\Bar R} ) \, {\Bar T}_{\Dot \b} ~+~  {\Bar R} \, ( {\Bar \de}_{\Dot\a} + \fracm12 {\Bar T}_{\Dot\a} ) \, {\Bar T}_{\Dot \b}  ~ \Big] \, {\cal M}_{\a\b}  \cr
& \qquad  +~ 2 \, C_{\Dot\a\Dot\b} \, {\Bar R} \, \Big[~ R ~+~ \frac 14  ( {\Bar\de}^{\Dot \g} {\Bar T}_{\Dot \g} )  ~ \Big] \, {\cal M}_{\a\b}  \cr
& -~ \frac 16 \,  C_{\Dot\a \Dot\b} \, \Big[~  R \, T_{\b} \, T_{\g} ~+~ \frac12 \, T_{\b} \, ( {\Bar\de}^{\Dot\d} {\Bar\de}_{\Dot\d} T_{\g} ) ~+~ \frac14 \, T_{\b} \, T_{\g} \, ( {\Bar\de}^{\Dot\d} {\Bar T}_{\Dot\d} )   ~\Big]  ~ {\cal M}_{\a}{}^{\g}  \cr 
& +~ \frac 16 \,  C_{\Dot\a \Dot\b} \, \Big[~ 2 \, (\de_{\a} R) \, T_{\g} ~+~ 2 \, R \, (\de_{\a} T_{\g} ) ~+~  \big(~ \de_{\a} {\Bar\de}^{\Dot \d} \, ( {\Bar\de}_{\Dot \d} ~+~ \fracm12 {\Bar T}_{\Dot \d} ) \, T_{\g} ~\big)   \cr
& \qquad +~ \fracm 12 ( {\Bar \de}^{\Dot\d} {\Bar T}_{\Dot \d} ) \, ( \de_{\a} T_{\g} )   ~\Big]  ~ {\cal M}_{\b}{}^{\g}  \cr 
& -~  C_{\a\b} \, \Big[~ ( {\Bar \de}_{\Dot \a} \de^{\g} 
G_{\d \Dot \b}  )  ~-~ \frac 12  \, {\Bar T}_{\Dot \b} \, ( \de^{\g} G_{\d \Dot \a}  )  ~\Big] \,  {\cal M}_{\g}{}^{\d}  \Big\}  \cr
& +~ \text{h. c.}
~~~~. 
 } \label{eq:uuuu}
\end{equation}
To our knowledge, the results in (3.14) mark the first time that a superspace description
of 4D, $\cal N$ = 1 Nordstr\" om supergravity has appeared in the literature.

To summarize, the limit of off-shell 4D, $\cal N$ = 1 superfield
supergravity where we {\em {only}} retain the conformal compensator provides a
superspace extension of the Nordstr\" om supergravitation theory that is discussed
in chapter three.  We will make a working assumption that such an approach
is universally applicable to all superspaces.  In particular, in the subsequent
chapters we will apply this assumption to superspaces whose bosonic
subspaces possess either eleven or ten dimensions.

\newpage
\section{Linearized Nordstr\" om Supergravity in 11D, $\mathcal{N}=1$ Supergeometry}

\indent
We begin our discussion by reviewing the work of \cite{crD11a,crD11b}
where it was shown that the entire structure of the torsions, curvatures, and
4-form field strengths could be written in terms of a single superfield
denoted by $W_{ {\un a} {\un b}  {\un c} {\un d}}$.   Using the conventions of
\cite{M2}, we can write
\be{
\eqalign{T_{\a\b} {}^{\un c} \, = & \, i (\g^{\un c})_{\a\b} ~~, ~~~
F_{\a\b {\un c}{\un d}} \,=\, \fracm 12 (\g_{{\un c} {\un d}})_{\a\b} ~~,
~~~
F_{{\un a} {\un b}  {\un c}  \d} = 0 ~~~, \cr
F_{\a\b\g\d} \,=&\, F_{\a\b\g {\un d}} \,=\, 0 ~~, ~~~ F_{{\un c} {\un d} 
{\un e} {\un f} }  ~=~ W_{{\un c} {\un d} {\un e} {\un f} } ~~~~~,  \cr 
T_{\a\b}{}^\g \,= & \, 0 ~~~, ~~~
T_{\a {\un b} } {}^{\un c}   \,= \, 0  ~~, \cr 
T_{\a {\un b} } {}^{\g} \, = \, & i \frac 1 {144}
(\g_{\un b}  {}^{{\un c} {\un d} {\un e} {\un f} } + 8 \d_{\un b} {}^{\un c} 
\g^{{\un d} {\un e} {\un f} } )_{\a}{}^{\g} W_{{\un c} {\un d} {\un e} {\un f} } 
~~~,\cr 
R_{\a\b {\un c} {\un d}} \, = \, & \frac 1 3 (\g^{{\un e} {\un f} })_{\a\b} 
W_{{\un c} {\un d} {\un e} {\un f} } ~+~  \frac 1 {72} (\g_{{\un c} 
{\un d}}{}^{{\un e} {\un f}  {\un g} {\un h} } )_{\a\b} 
W_{{\un e} {\un f}  {\un g} {\un h} } ~~~. } {~~~~~}
}\ee
In addition to the torsion and curvature supertensors, the formulation above
includes the 4-form supertensor, $F{}_{{\un A} {\un B} {\un C} {\un D} }$.  It should
be noted that these equations in (4.1) are the eleven dimensional analog of the
equations in (3.12).  In other words, the supergeometry in (4.1) is an ``on-shell'' 
supergeometry.  We must find a supergeometry consistent with the Norstr\" om 
theory as the analogs of (3.14).

We now wish to construct the linearized torsion and curvature supertensors with 
property that when all fermions are set to zero, the theory smoothly maps to
the linearization of the non-supersymmetrical theory described in chapter two. 

For this purpose we introduce eleven dimensional supergravity covariant 
derivatives linear in the infinitesimal conformal compensator $\Psi$ given by
\begin{align}
\nabla_{\alpha} ~=&~ {\rm D}_{\alpha} + \frac{1}{2}\Psi {\rm D}_{\alpha} +l_0({\rm 
D}_{\beta}\Psi)(\gamma^{\un{d}\un{e}})_{\alpha}^{\ \beta}{\cal M}_{\un{d}\un{e}}
~~~,  \\ 
\nabla_{\un{a}} ~=&~ \pa_{\un{a}}+\Psi\pa_{\un{a}}+il_1(\gamma_{\un{a}})^{\alpha
\beta}({\rm D}_{\alpha}\Psi){\rm D}_{\beta}+l_2(\pa_{\un{c}}\Psi){\cal M}_{\un{a}}^{\ 
\un{c}}\nonumber  ~~~,  \\
&+il_3(\gamma_{\un{a}}^{\ \un{d}\un{e}})^{\alpha\beta}({\rm D}_{\alpha}{\rm D}_{
\beta}\Psi){\cal M}_{\un{d}\un{e}} ~~~,
\end{align}
where the ``bare'' superderivative operators ${\rm D}_{\alpha}$ satisfy
\begin{equation}
\{{\rm D}_{\alpha},{\rm D}_{\beta}\} = i(\gamma^{\un{a}})_{\alpha\beta}\pa_{\un{a}}
~~~,
\end{equation}
and the torsion tensors and Riemann curvature tensors can be obtained via 
\begin{equation}
\begin{split}
[\nabla_{\alpha},\nabla_{\beta}\} ~=&~ T_{\alpha\beta}{}^{\un{c}}\nabla_{\un{c}}+
T_{\alpha\beta}{}^{\gamma}\nabla_{\gamma} + \frac{1}{2}R_{\alpha\beta\un{
d}}{}^{\un{e}} \mathcal{M}_{\un{e}}{}^{\un{d}} ~~~,  \\
[\nabla_{\alpha},\nabla_{\un{b}}\} ~=&~ T_{\alpha\un{b}}{}^{\un{c}}\nabla_{\un{c}}+
T_{\alpha\un{b}}{}^{\gamma}\nabla_{\gamma} + \frac{1}{2}R_{\alpha\un{b}\un{
d}}{}^{\un{e}} \mathcal{M}_{\un{e}}{}^{\un{d}} ~~~,  \\
[\nabla_{\un{a}},\nabla_{\un{b}}\} ~=&~ T_{\un{a}\un{b}}{}^{\un{c}}\nabla_{\un{c}}+
T_{\un{a}\un{b}}{}^{\gamma}\nabla_{\gamma} + \frac{1}{2}R_{\un{a}\un{b}\un{
d}}{}^{\un{e}} \mathcal{M}_{\un{e}}{}^{\un{d}} ~~~.
\end{split}
\end{equation}
The commutation relations of the operators with the 11D Lorentz generators satisfy
\begin{align}
&[ {\cal M}_{\un{a}\un{b}},{\rm D}_{\alpha} \} ~=~ \frac{1}{2}(\gamma_{\un{a}\un{b}}
)_{\alpha}^{\ \beta}{\rm D}_{\beta}   ~~~, \\
&\big[ {\cal M}_{\un{a}\un{b}},[{\cal M}_{\un{c}\un{d}},{\rm D}_{\alpha}\}\big\} + \big[{\cal M}_{\un{c}\un{d}},[{\rm D}_{\alpha},{\cal M}_{\un{a}\un{b}}\}\big\} + \big[{\rm D}_{\alpha},[{\cal 
M}_{\un{a}\un{b}},{\cal M}_{\un{c}\un{d}}\}\big\} ~=~ 0 ~~~,
\end{align}
in addition to the relations seen in (2.1) and (2.2).

By imposing the constraints
\begin{align}
T_{\un{a}\un{b}}^{\ \ \un{c}}&~=~ 0  ~~~,~~~
T_{\alpha\beta}^{\ \ \ \un{c}} ~=~ i(\gamma^{\un{c}})_{\alpha\beta}  ~~~,
\end{align}
we obtain the following parameterization results:
\begin{align}
l_0 ~=~ \frac{1}{10}  ~~~,~~~
l_1 ~=~ \frac{1}{4}  ~~~,~~~
l_2 ~=~ -1 ~~~,~~~
l_3 ~=~ 0  ~~~.
\end{align}
In turn these lead to a set of results that express the torsion and curvature tensors
solely in terms of $\Psi$ and its derivatives.  We give these in the following two
subsections.  

For the components of the torsion we find the results seen in (4.10)
- (4.15).
\begin{align}
{~~~~~~~~~~}
T_{\alpha\beta}^{\ \ \un{c}}  ~=~& i(\gamma^{\un{c}})_{\alpha\beta} ~~~,  &&\\
T_{\alpha\beta}^{\ \ \gamma} ~=~& \frac{3}{40}(\gamma^{[2]})_{\alpha\beta}(\gamma_{[2]})^{\gamma\delta} ({\rm D}_{\delta}\Psi)  ~~~, &&\\
T_{\alpha\un{b}}^{\ \ \un{c}} ~=~&   \frac{3}{4} \delta_{\un{b}}^{\ \un{c}} ({\rm D}_{\alpha}\Psi) + \frac{9}{20}(\gamma_{\un{b}}{}^{\un{c}})_{\alpha}^{\ \beta}({\rm D}_{\beta} \Psi)  ~~~, &&\\
T_{\alpha\un{b}}^{\ \ \gamma} ~=~& i\frac{1}{128} \Big[ - (\gamma_{\un{b}})_{\alpha}^{\ \gamma} C^{\delta\epsilon} + \frac{1}{2} (\gamma^{[2]})_{\alpha}^{\ \gamma} (\gamma_{\un{b}[2]})^{\delta\epsilon} - \frac{1}{3!} (\gamma_{\un{b}[3]})_{\alpha}^{\ \gamma} (\gamma^{[3]})^{\delta\epsilon}  + \frac{1}{3!} (\gamma^{[3]})_{\alpha}^{\ \gamma} (\gamma_{\un{b}[3]})^{\delta\epsilon}  \nonumber\\
& \qquad   - \frac{1}{4!} (\gamma_{\un{b}[4]})_{\alpha}^{\ \gamma} (\gamma^{[4]})^{\delta\epsilon}   \Big] ({\rm 
D}_{\delta}{\rm D}_{\epsilon}\Psi) + \frac{1}{8} \delta_{\alpha}^{\ \gamma} (\pa_{\un{b}}\Psi) + \frac{3}{8} (\gamma_{\un{b}}{}^{\un{c}})_{\alpha}^{\ \gamma} (\pa_{\un{c}}\Psi) ~~~,  &&\\
T_{\un{a}\un{b}}^{\ \ \un{c}} ~=~& 0 ~~~,  &&\\
T_{\un{a}\un{b}}^{\ \ \gamma} ~=~&  -i\frac{1}{4}(\gamma_{[\un{a}})^{\gamma\delta}
(\pa_{\un{b}]}{\rm D}_{\delta}\Psi) ~~~.
\end{align}
For the components of the curvature we find the results seen in (4.16)
- (4.18).
\begin{align}
{~~~~~~~~~~}
R_{\alpha\beta}^{\ \ \ \un{d}\un{e}} ~=~& \frac{1}{80}  \Big[  (\gamma^{\un{d}\un{e}})_{\alpha\beta} C^{\gamma\delta} + (\gamma_{[1]})_{\alpha\beta}(\gamma^{[1]\un{d}\un{e}})^{\gamma\delta} - \frac{1}{2} (\gamma_{[2]})_{\alpha\beta} (\gamma^{[2]\un{d}\un{e}})^{\gamma\delta}   - \frac{1}{3!} (\gamma^{\un{d}\un{e}[3]})_{\alpha\beta}(\gamma_{[3]})^{\gamma\delta} \nonumber \\
& \qquad   + \frac{1}{5!4!} \epsilon^{\un{d}\un{e}[5][4]} (\gamma_{[5]})_{\alpha\beta} (\gamma_{[4]})^{\gamma\delta}     \Big] ({\rm D}_{\gamma}{\rm 
D}_{\delta}\Psi) ~~~,  &&\\
R_{\alpha\un{b}}^{\ \ \ \un{d}\un{e}} ~=~& -(\partial^{[\un{d}}{\rm D}_{\alpha}\Psi)
\delta_{\un{b}}^{\ \un{e}]}+\frac{1}{5}(\gamma^{\un{d}\un{e}})_{\alpha}^{\ \delta}
(\pa_{\un{b}}{\rm D}_{\delta}\Psi) ~~~, &&\\
R_{\un{a}\un{b}}^{\ \ \ \un{d}\un{e}} ~=~& -(\pa_{[\un{a}}\partial^{[\un{d}}\Psi)\delta_{
\un{b}]}^{\ \un{e}]} ~~~.
\end{align}
In reaching (4.10) - (4.18), we used the Fierz identities (A.24) - (A.27) listed in Appendix A. 

It is the last equation that ensures that we have reached our goal.  Namely, the choice of
constraints in (4.9) has led to a linearized super Riemann curvature tensor expressed 
solely in terms of an infinitesimal superfield $\Psi$ that has the exact form of the first
term in the non-supersymmetrical Riemann curvature tensor given in (2.8).  Recall that
the supersymmetrical theory here is linearized, so to make a proper comparison
to the bosonic theory, that should also be linearized.  When this is done, there is a matching of the terms.

We should note the work in \cite{M2} also constructs a fully non-linear 11D supergeometry
in terms of a finite scalar compensator.  However, its linearization is different from the
one obtained here.  In the next three chapters, we will obtain new and never before
presented results of this nature for the 10D, $\cal N$ = 2A, 10D, $\cal N$ = 2B, and 
10D, $\cal N$ = 1 supergeometries that possess the purely bosonic linearized results
as in the linearization of (2.8). The Fierz identities used for simplifying the torsions and curvatures are listed in Appendix B.

\newpage
\section{Linearized Nordstr\" om Supergravity in 10D, $\cal N$ = 1 Supergeometry}

We begin this discussion by pointing out the on-shell description of 10D, $\cal N$ = 1
superspace supergravity.  A set of torsion and curvature supertensors can be written
in the form 
\be  \eqalign {  {~~~~~~}
T_{\a\b}{}^{\un c}~&=~ i\,(\s^{\un c}){}_{\a\b}~~,~~~
T_{\a\b}{}^{\g}~=-~\fracm12\,\sqrt{\fracm12}\,
\Big[\d_{(\a}{}^{\g}\,\d_{\b)}{}^{\e} ~+~(\s^{\un a}){}_{\a\b}
(\s_{\un a}){}^{\g\,\e}\,\Big]\,\c_{\e}~~,~~~ T_{\a\un b}{}^{\un c}~=~0~~,\cr
T_{\a\un{b}}{}^{\g}~&=~ -{\fracm{1}{24}}(\s_{\un
b}\,\s^{ {\un c}{\un d}{\un e}})_\a{}^\g\,\Big[\,e^{\Phi}\,H_{ {\un c}{\un d}{\un e}}
-i\,\fracm18\,(\,\c\,\s_{ {\un c}{\un d}{\un e}}\,\c)
\, \, \Big]\cr
&{~~~~~}  -{\fracm{1}{48}}(\s^{ {\un c}{\un d}{\un e}}\,\s_{\un b})_\a{}^\g\,\Big[
\,e^{\Phi}\,H_{ {\un c}{\un d}{\un e}}-i\,{\fracm{1}{16}}\,(\,\c\,
\s_{ {\un c}{\un d}{\un e}}\,\c)  \, 
\Big]~~,\cr
R_{\a\b\un{d}\un{e}}~&=~ -i\,\fracm14\,(\s^{\un c}){}_{\a\b}\,
\Big[\,3e^{-\Phi}\,H_{\un{c}\un{d}\un{e}}-i\,{\fracm{5}{16}}\,(\,\c\,
\s_{ {\un c}{\un d}{\un e}}\,\c)   \, \Big]\cr
&{~~~~~}-i\,{\fracm{1}{24}}\,(\s^{ {\un a}{\un b}{\un c}}{}_{\un{d}\un{e}})_{\a\b}
\, \Big[\,e^{-\Phi}\,H_{ {\un a}{\un b}{\un c}}-i\,{\fracm{3}{16}}\,
(\,\c\,\s_{{\un a}{\un b}{\un c}}\,\c)  \, \Big]~~,\cr
R_{\a\un{c}\un{d}\un{e}}~&=~ -i\,\fracm12\Big[\,(\s_{\un c})_{\a\g}\,T_{\un {d}\un{e}}{}^{\g}-(\s_{\un d})_{\a\g}\,T_{\un{
e}\un{c}}{}^{\g}\,-(\s_{\un e})_{\a\g}\,T_{\un{c}\un{d}}{}^{\g}
\Big]~~.} 
\ee 
as was noted in the work of \cite{FX1,Gates:1986tj}.  In these expression $H_{ {\un a}{\un b}{\un c}}$
refers to the supercovariantized field strength of a two-form $B_{ {\un a}{\un b}}$.  The
results in (5.1) are the 10D, $\cal N$ = 1 analogs of the results in (3.12) for the 4D, $\cal N$ = 1 superspace geometry.  That is the component fields embedded in this supergeometry
must obey a set of mass-shell conditions.  To release these conditions, one must find
the 10D, $\cal N$ = 1 analogs of the equations in (3.9) and (3.10).  However, as our goal once
more is to find a supergeometry consistent with the Norstr\" om theory, we seek the
analogs of (3.14).

The covariant derivatives linear in the conformal compensator $\Psi$ are given by
\begin{flalign}  {~~~~~~~~~~}
\nabla_{\alpha} &= {\rm D}_{\alpha}+l_0\Psi {\rm D}_{\alpha}+l_1(\sigma^{\un{a}\un{b}})_{
\alpha}^{\ \beta}({\rm D}_{\beta}\Psi){\cal M}_{\un{a}\un{b}}  ~~~, &&\\
\nabla_{\un{a}} &= \pa_{\un{a}} + l_2\Psi\pa_{\un{a}} + il_3(\sigma_{\un{a}})^{\alpha\beta}
({\rm D}_{\alpha}\Psi){\rm D}_{\beta} +l_4 (\partial_{\un{c}}\Psi){\cal M}_{\un{a}}^{\ \un{c}} +il_5(
\sigma_{\un{a}}^{\ \un{d}\un{e}})^{\gamma\delta}({\rm D}_{\gamma}{\rm D}_{\delta}\Psi)
{\cal M}_{\un{d}\un{e}} ~~~,
\end{flalign}
and similar to the case of the eleven dimensional theory, here we have
\begin{equation}
\{{\rm D}_{\alpha},{\rm D}_{\beta}\} = i(\sigma^{\un{a}})_{\alpha\beta}\pa_{\un{a}}
~~~.
\end{equation}
The commutation relations of Poincare generators in 10D
\begin{align}
&[{\cal M}_{\un{a}\un{b}},{\rm D}_{\alpha}\} = \frac{1}{2}(\sigma_{\un{a}\un{b}})_{\alpha}^{\ 
\beta}{\rm D}_{\beta} ~~~,
\end{align}
is similar to the eleven dimensional case.  Also the equation in (4.7) is valid in all
ten dimensional theories.  There will be some slight modifications for the dotted 
and barred spinor indices in type IIA and IIB supergravity, respectively.

By adoption of the constraints
\begin{align}
T_{\un{a}\un{b}}^{\ \ \un{c}}&=0 ~~~,~~~
T_{\alpha\beta}^{\ \ \ \un{c}}=i(\sigma^{\un{c}})_{\alpha
\beta}  ~~~,
\end{align}
we obtain the following parameterization results:
\begin{align}
l_0 &= \frac{1}{2} ~~,~~
l_1 = \frac{1}{10} ~~,~~
l_2 = 1 ~~,~~
l_3 = -\frac{2}{5} ~~,~~
l_4 =  -1 ~~,~~
l_5 =  0 ~~.
\end{align}

As the consequence of this choice of parameters, we find the torsion supertensors given in (5.8) - (5.13).
\begin{align} {~~~~~~~~~~} 
T_{\alpha\beta}^{\ \ \un{c}} ~=~ & i(\sigma^{\un{c}})_{\alpha\beta} ~~~, &&\\
T_{\alpha\beta}^{\ \ \gamma} ~=~ & 0 ~~~, &&\\
T_{\alpha\un{b}}^{\ \ \un{c}} ~=~ & \frac{3}{5} \left[
\delta_{\un{b}}^{\ \un{c}}\delta_{\alpha}^{\ \delta} + (\sigma_{\un{b}}^{\ \un{c}})_{\alpha}^{\ \delta
}\right]({\rm D}_{\delta}\Psi) ~~~, &&\\
T_{\alpha\un{b}}^{\ \ \gamma} ~=~ & i\frac{1}{80} \left[-(\sigma^{[2]})_{\alpha}^{\ 
\gamma}(\sigma_{\un{b}[2]})^{\beta\delta} +  \frac{1}{3} (\sigma_{\un{b}[3]})_{\alpha}^{\ \gamma}(\sigma^{[3]})^{\beta\delta}\right]({\rm D}_{\beta}{\rm D}_{\delta}\Psi) \nonumber\\
& - \frac{3}{10} \delta_{\alpha}^{\ \gamma} (\pa_{\un{b}}\Psi) + \frac{3}{10} (\sigma_{\un{b}}^{\ \un{c}})_{\alpha}^{\ \gamma} (\pa_{\un{c}}\Psi)   ~~~, &&\\
T_{\un{a}\un{b}}^{\ \ \un{c}} ~=~ & 0 ~~~,  &&\\
T_{\un{a}\un{b}}^{\ \ \gamma} ~=~ & i\frac{2}{5}(\sigma_{[\un{a
}})^{\gamma\delta}(\pa_{\un{b}]}{\rm D}_{\delta}\Psi) ~~~.
\end{align}
For the components of the curvatures, we find the results seen in (5.14) - (5.16).
\begin{align} {~~~~~~~~~~}
R_{\alpha\beta}^{\ \ \ \un{d}\un{e}} ~=~ & -i\frac{6}{5}(\sigma^{[\un{d}})_{\alpha\beta}
(\partial^{\un{e}]}\Psi) - \frac{1}{40}\left[\frac{1}{3!}(\sigma^{\un{d}\un{e}[3]})_{\alpha\beta}(\sigma_{[3]})^{
\gamma\delta}+(\sigma^{\un{a}})_{\alpha\beta}(\sigma_{\un{a}}^{\ \un{d}\un{e}})^{\gamma\delta} \right] ({\rm D}_{\gamma}{\rm D}_{\delta}\Psi)~~~,  &&\\
R_{\alpha\un{b}}^{\ \ \ \un{d}\un{e}} ~=~ & -({\rm D}_{\alpha}\partial^{[\un{d}}
\Psi)\delta_{\un{b}}^{\ \un{e}]}+\frac{1}{5}(\sigma^{\un{d}\un{e}})_{\alpha
}^{\ \gamma}(\pa_{\un{b}}{\rm D}_{\gamma}\Psi)  ~~~,  &&\\
R_{\un{a}\un{b}}^{\ \ \ \un{d}\un{e}} ~=~ & -(\pa_{[\un{a}}\partial^{[\un{d}}
\Psi)\delta_{\un{b}]}^{\ \un{e}]}  ~~~.
\end{align}

It has long been suggested \cite{HNVP} that a superfield with the structure of 
$G_{{\un a} {\un b} {\un c}}$ should appear in the off-shell structure of 10D, $\cal N$ 
= 1 supergeometry and that it was related by a superdifferential operator to an
underlying unconstrained prepotential $V_{{\un a} {\un b} {\un c}}$ analogous to 
$H^{\un m}$ that appears in 4D, $\cal N$ = 1 supergravity.  However, there are
reasons to believe \cite{FX1,Gates:1986tj} that $V_{{\un a} {\un b} {\un c}}$ must be related to an
even more fundamental spinorial prepotential ${\Psi}_{{\un a} {\un b}}{}^{\a}$.
In the equations of (5.11) and (5.14) the superfield $G_{{\un a} {\un b} {\un c}}\equiv (\sigma_{\un{a}\un{b}\un{c}})^{\gamma\delta} ({\rm D}_{\gamma}{\rm D}_{\delta}\Psi)$ has precisely the structure suggested in the work by Howe, Nicolai, and Van Proeyen.

\newpage
\section{Linearized Nordstr\" om Supergravity in 10D, $\cal N$ = 2A Supergeometry}

We repeat the discussions as seen in the previous two chapters with a beginning
of the on-shell description of 10D, $\cal N$ = 2A superspace supergravity.  A set 
of torsion and curvature supertensors can be written in the form
\be
\eqalign{
T_{\a\Dot\b}{}^{\un c} ~&=~T_{\Dot\a\Dot\b}{}^\g ~=~T_{\a\Dot\b}{}^\g ~=~
T_{\Dot\a \b}{}^{\Dot\g}~=~T_{\a\b}{}^{\Dot\g}~=~0~~,\cr
T_{\a \un b}{}^{\un c} ~&=~T_{\Dot\a \un b}{}^{\un c}~=~ T_{\un a\un b}{}^{
\un c}~ =~0~~~,\cr
T_{\a\b}{}^{\un c} ~&=~ i (\s^{\un c} )_{\a\b}~~~,~~~~T_{\Dot\a \Dot
\b}{}^{\un c}~ = ~i(\s^{\un c})_{\Dot
\a \Dot\b}~~~,\cr
T_{\a\b}{}^{\g} ~&=~ \Big[\d_{(\a}{}^\g \d_{\b)}{}^\d~+~(\s^{\un a})_{\a\b}
(\s_{\un a}){}^{\g\d}\Big] \chi_{\d}~~,\cr
T_{\Dot\a \Dot\b}{}^{\Dot\g} ~&=~ \Big[\d_{( \Dot\a}{}^{\Dot\g} \d_{\Dot\b)}{}^{\Dot\d}
~+~(\s^{\un a} )_{\Dot\a \Dot\b}(\s_{\un a} )^{\Dot\g \Dot\d}\Big]
\chi_{\Dot\d}~~~,\cr
T_{\a \un b}{}^{\g} ~&=~ -\frac18 (\s^{\un d\un e})_{\a}{}^{\g} H_{\un b\un d \un e}~~~,~~~~
C_{\a\Dot\a}C^{\g\Dot\g}T_{\Dot\g \un b}{}^{\Dot\a}~=~- \frac18 (\s^{\un d\un e})_{\a}{}^{\g} G_{\un b\un d\un e}~~~, \cr
C_{\g\Dot\g} T_{\a \un b}{}^{\Dot\g} ~&=~ \frac1{16}(\s_{\un b})_{\a\d} \Big[
(\s^{[2]})_{\g}{}^{\d} K_{[2]}~-~\fracm 1{12}(\s^{[4]})_{\g}
{}^{\d} D_{[4]} \Big]~~~, \cr
C^{\a\Dot\a}T_{\Dot\a \un b}{}^{\g} ~&=~ -\frac1{16}(\s_{\un b} )^{\a\d} \Big[
(\s^{[2]})_{\d} {}^{\g} K_{[2]}~-~\fracm 1{12}(\s^{[4]})_{\d}
{}^{\g} D_{[4]} \Big]~~~, } 
\ee
as was noted in the work of \cite{TP2}.  In these expressions $H_{ {\un a}{\un b}{\un c}}$ refers to the supercovariantized field strengths of a two-form $B_{ {\un a}{\un b}}$ gauge
field, and
\be\eqalign{
K_{\un a\un b}&=~e^{-\F}F_{\un a\un b}~-~\chi_{\a}(\s_{\un a\un b})_{\b}
{}^{\a} \chi^{\b}~~~,\cr
D_{[4]}&=~2e^{-\F}{\tilde F}_{[4]}~+~\chi_{\a}(\s_{[4]})_{\b}
{}^{\a} \chi^{\b}~~~.}\ee
with $F_{\un a\un b}$ and ${\tilde F}_{[4]}$ denoting supercovariantized field
strength for a gauge 1-form and a gauge 3-form respectively.  The results in (6.1) are the 10D, $\cal N$ = 2A analogs of the results in (3.12) for the 4D, $\cal N$ 
= 1 superspace geometry.  Those are the component fields embedded in this supergeometry
must obey a set of mass-shell conditions.  To release these conditions, one must find
the 10D, $\cal N$ = 2A analogs of the equations in (3.9) and (3.10).  Again the goal must be
to find a supergeometry consistent with the Norstr\" om theory, we seek the
analogs of (3.14).  In analogy with the 3-form gauge field sector of 11D, $\cal N$
= 1 supergravity the gauge fields components are:
\be
\eqalign{
{F}_{\a\Dot\b}&=~C_{\a\Dot\b}e^{\F}~~~,~~~~~~~~~~~~~~~
{F}_{\a\b}~=~{F}_{\Dot\a \Dot\b}~ =~0~~~,~\cr
{F}_{\un c\a}&=~ie^{\F}(\s_{\un c} )_{\a\b}\chi^{\b}~~~,~~
~~~~~~{F}_{\un c\Dot\a}~=~
iC_{\a\Dot\a}e^{\F}(\s_{\un c} )^{\a\b}\chi_{\b}~~,\cr
G_{\a\b\g}&=~G_{\un a\b\Dot\g}~=~G_{\un a\un b\g}~=~G_{\un a\un b\Dot\g}~=~0
~~~,\cr
G_{\un c\a\b}&=~i(\s_{\un c} )_{\a\b}~~~,~~~~~~~~G_{\un c\Dot\a \Dot\b}~=~- i(\s_{\un c} )_{\Dot\a 
\Dot\b}~~~,\cr
{\tilde F}_{\a\b\g \un d}&=~{\tilde F}_{\a\b \un c\un d}~=~
{\tilde F}_{\Dot\a \Dot\b \un c\un d}~=~0~~,~\cr
{\tilde F}_{\a\Dot\b \un c\un d}&=~e^{\F}(\s_{\un c\un d})_{\a} 
{}^{\b}C_{\b\Dot\b}~~~,~\cr
{\tilde F}_{\a \un b\un c\un d}&=~-ie^{\F}(\s_{\un b\un c\un
d})_{\a\b}\chi^{\b}~~~,~~~
~~~~{\tilde F}_{\Dot\a \un b\un c\un d}~=~
iC_{\a\Dot\a}e^{\F}(\s_{\un b\un c\un d})^{\a\b}\chi_{\b}~~~.}\ee
\noindent 
$\F$ denotes a dilaton superfield, and $\chi_{\a}$ is its partner dilatino.  All of
the equations in (6.1) - (6.3) describe the on-shell 10D, $\cal N$ = 2A theory,
i.e. these are the analogs of (3.12).

The covariant derivatives linear in the conformal compensator $\Psi$ are given by
\begin{flalign}  {~~~~~~~~~~}
\nabla_{\alpha} ~= &~{\rm D}_{\alpha} + \frac{1}{2}\Psi {\rm D}_{\alpha}+l_0(\sigma^{\un{
a}\un{b}})_{\alpha}^{\ \beta}({\rm D}_{\beta}\Psi){\cal M}_{\un{a}\un{b}}&&\\
\nabla_{\Dot{\alpha}} ~= &~{\rm D}_{\Dot{\alpha}} + \frac{1}{2}\Psi {\rm D}_{\Dot{\alpha}}
+l_0(\sigma ^{\un{a}\un{b}})_{\Dot{\alpha}}^{\ \Dot{\beta}}({\rm D}_{\Dot{\beta}}\Psi){\cal 
M}_{\un{a}\un{b}}&&\\
\nabla_{\un{a}}~= &~\pa_{\un{a}}+l_1\Psi\pa_{\un{a}}+il_2(\sigma_{\un{a}})^{\delta 
\gamma}({\rm D}_{\delta}\Psi){\rm D}_{\gamma}+il_3(\sigma_{\un{a}})^{\Dot{\delta}
\Dot{ \gamma}}({\rm D}_{\Dot{\delta}}\Psi){\rm D}_{\Dot{\gamma}}+l_4(\pa_{\un{c}}
\Psi){\cal M}_{\un{a}}^{\ \un{c}}\nonumber&&\\
&+il_5(\sigma_{\un{a}}^{\ \un{c}\un{d}})^{\gamma\delta}({\rm D}_{\gamma}{\rm D
}_{\delta}\Psi){\cal M}_{\un{c}\un{d}}+il_6(\sigma_{\un{a}}^{\ \un{c}\un{d}})^{\Dot{
\gamma}\Dot{\delta}}({\rm D}_{\Dot{\gamma}}{\rm D}_{\Dot{\delta}}\Psi){\cal M}_{
\un{c}\un{d}}
\end{flalign}
where the Type IIA supersymmetry algebra 
\begin{align}
\{{\rm D}_{\alpha},{\rm D}_{\beta}\} ~&=~ i(\sigma^{\un{a}})_{\alpha\beta}\pa_{\un{a}}~~,~~
\{{\rm D}_{\Dot{\alpha}},{\rm D}_{\Dot{\beta}}\} ~=~ i(\sigma^{\un{a}})_{\Dot{\alpha}\Dot{
\beta}}\pa_{\un{a}} ~~,~~
\{{\rm D}_{\alpha},{\rm D}_{\Dot{\beta}}\} ~=~ 0 ~~
\end{align}
is satisfied by the bare derivative operators.

By adoption the constraints
\begin{align}
T_{{\un a} {\un b}}^{\ \ \un{c}}&=0 ~~~,~~~~
T_{\alpha\beta}^{\ \ \un{c}}=i(\sigma^{\un{c}})_{\alpha\beta}~~~,~~~~
T_{\Dot{\alpha}\Dot{\beta}}^{\ \ \un{c}}=i(\sigma^{\un{c}})_{\Dot{\alpha}\Dot{\beta}}
~~~,
\end{align}
we obtain the following paramaterization values: 
\begin{align}
l_0 &= \frac{1}{10} ~~,~~
l_1 = 1  ~~,~~
l_2 = l_3 = -\frac{1}{5} ~~,~~
l_4 = -1 ~~,~~
l_5 = l_6 = 0 ~~.
\end{align}

As the consequence of this choice of parameters, we find the torsion supertensors given in (6.10)
- (6.27). 
\begin{align} {~~~~~~~~~~} 
T_{\alpha\beta}^{\ \ \un{c}} ~=~ & i(\sigma^{\un{c}})_{\alpha\beta} ~~~, &&\\
T_{\alpha\beta}^{\ \ \gamma} ~=~ & \frac{1}{5}(\sigma^{\un a})_{\alpha\beta}(\sigma_{\un a})^{\gamma\delta}({\rm D}_{\delta}\Psi) ~~~, &&\\
T_{\alpha\beta}^{\ \ \Dot\gamma} ~=~ & -\frac{1}{5}(\sigma^{\un a})_{\alpha\beta}(\sigma_{\un a})^{\Dot\gamma\Dot\delta}({\rm D}_{\Dot\delta}\Psi) ~~~, &&\\
{~~~~~~~~~~} T_{\Dot\alpha\Dot\beta}^{\ \ \un{c}} ~=~ & i(\sigma^{\un{c}})_{\Dot\alpha\Dot\beta} ~~~, &&\\
T_{\Dot\alpha\Dot\beta}^{\ \ \gamma} ~=~ & -\frac{1}{5}(\sigma^{\un a})_{\Dot\alpha\Dot\beta}(\sigma_{\un a})^{\gamma\delta}({\rm D}_{\delta}\Psi) ~~~, &&\\
T_{\Dot\alpha\Dot\beta}^{\ \ \Dot\gamma} ~=~ & \frac{1}{5}(\sigma^{\un a})_{\Dot\alpha\Dot\beta}(\sigma_{\un a})^{\Dot\gamma\Dot\delta}({\rm D}_{\Dot\delta}\Psi) ~~~, &&\\
T_{\alpha\Dot\beta}^{\ \ \un c} ~=~ & 0 ~~~, &&\\
T_{\alpha\Dot\beta}^{\ \ \gamma} ~=~ & \frac{1}{2}\left[ \delta_{\alpha}^{\ \gamma} ({\rm D}_{\Dot\beta}\Psi) + \frac{1}{10} (\sigma^{\un a \un b})_{\alpha}^{\ \gamma}(\sigma_{\un a\un b})_{\Dot\beta}^{\ \Dot\delta}({\rm D}_{\Dot\delta}\Psi)\right] ~~~, &&\\
T_{\alpha\Dot\beta}^{\ \ \Dot\gamma} ~=~ & \frac{1}{2}\left[ \delta_{\Dot\beta}^{\ \Dot\gamma} ({\rm D}_{\alpha}\Psi) + \frac{1}{10} (\sigma^{\un a\un b})_{\Dot\beta}^{\ \Dot\gamma} (\sigma_{\un a \un b})_{\alpha}^{\ \delta} ({\rm D}_{\delta}\Psi)\right] ~~~, &&\\
T_{\alpha\un{b}}^{\ \ \un{c}} ~=~ & \frac{4}{5}\delta_{\un{b}}^{\ \un{c}}({\rm D}_{\alpha}\Psi)+\frac{2}{5}(\sigma_{\un{b}}^{\ \un{c}})_{\alpha}^{\ \delta
}({\rm D}_{\delta}\Psi) ~~~, &&\\
T_{\alpha\un{b}}^{\ \ \gamma} ~=~ & i\frac{1}{80}\left[-\frac{1}{2}(\sigma^{[2]})_{\alpha}^{\ 
\gamma}(\sigma_{\un{b}[2]})^{\beta\delta}+ \frac{1}{3!}(\sigma_{\un{b}[3]})_{\alpha}^{\ \gamma}(\sigma^{
[3]})^{\beta\delta}\right]({\rm D}_{\beta}{\rm D}_{\delta}\Psi) \nonumber\\
& - \frac{2}{5}\delta_{\alpha}^{\ \gamma}(\pa_{
\un{b}}\Psi) + \frac{2}{5} (\sigma_{\un{b}}^{\ \un{c}})_{\alpha}^{\ \gamma} (\pa_{\un{
c}}\Psi)  ~~~, &&\\
T_{\alpha\un{b}}^{\ \ \Dot\gamma} ~=~ \nonumber& -i\frac{1}{80}\Big[(\sigma_{\un{b}})_{\alpha}^{\ 
\Dot\gamma}C^{\beta\Dot\delta}-(\sigma^{\un c})_{\alpha}^{\ \Dot\gamma}(\sigma_{\un b\un c})^{\beta\Dot\delta}-\frac{1}{3!}(\sigma^{[3]})_{\alpha}^{\ \Dot\gamma}(\sigma_{\un b[3]})^{\beta\Dot\delta}+\frac{1}{2}(\sigma_{\un b[2]})_{\alpha}^{\ \Dot\gamma}(\sigma^{[2]})^{\beta\Dot\delta}&&\\&+\frac{1}{4!}(\sigma_{\un b[4]})_{\alpha}^{\ \Dot\gamma}(\sigma^{[4]})^{\beta\Dot\delta}\Big]
({\rm D}_{\beta}{\rm D}_{\Dot\delta}\Psi)  ~~~, &&\\
T_{\Dot\alpha\un{b}}^{\ \ \un{c}} ~=~ & \frac{4}{5} \delta_{\un{b}}^{\ \un{c}} ({\rm D}_{\Dot\alpha}\Psi) + \frac{2}{5} (\sigma_{\un{b}}^{\ \un{c}})_{\Dot\alpha}^{\ \Dot\delta
}({\rm D}_{\Dot\delta}\Psi) ~~~, &&\\
T_{\Dot\alpha\un{b}}^{\ \ \gamma} ~=~ \nonumber& -i\frac{1}{80}\Big[(\sigma_{\un{b}})_{\Dot\alpha}^{\ 
\gamma}C^{\delta\Dot\beta}+(\sigma^{\un c})_{\Dot\alpha}^{\ \gamma}(\sigma_{\un b\un c})^{\delta\Dot\beta}-\frac{1}{3!}(\sigma^{[3]})_{\Dot\alpha}^{\ \gamma}(\sigma_{\un b[3]})^{\delta\Dot\beta}-\frac{1}{2}(\sigma_{\un b[2]})_{\Dot\alpha}^{\ \gamma}(\sigma^{[2]})^{\delta\Dot\beta}&&\\&+\frac{1}{4!}(\sigma_{\un b[4]})_{\Dot\alpha}^{\ \gamma}(\sigma^{[4]})^{\delta\Dot\beta}\Big]({\rm D}_{\delta}{\rm D}_{\Dot\beta}\Psi)  ~~~, &&\\
T_{\Dot\alpha\un{b}}^{\ \ \Dot\gamma} ~=~ & i\frac{1}{80}\left[-\frac{1}{2}(\sigma^{[2]})_{\Dot\alpha}^{\ 
\Dot\gamma}(\sigma_{\un{b}[2]})^{\Dot\beta\Dot\delta}+ \frac{1}{3!}(\sigma_{\un{b}[3]})_{\Dot\alpha}^{\ \Dot\gamma}(\sigma^{[3]})^{\Dot\beta\Dot\delta}\right]({\rm D}_{\Dot\beta}{\rm D}_{\Dot\delta}\Psi) \nonumber\\
&  - \frac{2}{5} \delta_{\Dot\alpha}^{\ \Dot\gamma} (\pa_{\un{b}}\Psi) + \frac{2}{5} (\sigma_{\un{b}}^{\ \un{c}})_{\Dot\alpha}^{\ \Dot\gamma} (\pa_{\un{c}}\Psi)   ~~~, &&\\
T_{\un{a}\un{b}}^{\ \ \un{c}} ~=~ & 0 ~~~,  &&\\
T_{\un{a}\un{b}}^{\ \ \gamma} ~=~ & i\frac{1}{5}(\sigma_{[\un{a
}})^{\gamma\delta}(\pa_{\un{b}]}{\rm D}_{\delta}\Psi) ~~~, &&\\
T_{\un{a}\un{b}}^{\ \ \Dot\gamma} ~=~ & i\frac{1}{5}(\sigma_{[\un{a
}})^{\Dot\gamma\Dot\delta}(\pa_{\un{b}]}{\rm D}_{\Dot\delta}\Psi) ~~~.
\end{align}

For the components of the curvatures, we find the results seen in (6.28) - (6.33).
\begin{align} {~~~~~~~~~~} 
R_{\alpha\beta}^{\ \ \ \un{d}\un{e}} ~=~ & -i\frac{6}{5}(\sigma^{[\un{d}})_{\alpha\beta}
(\partial^{\un{e}]}\Psi) - \frac{1}{40} \left[\frac{1}{3!}(\sigma^{\un{d}\un{e}[3]})_{\alpha\beta}(\sigma_{[3]})^{\gamma\delta}+(\sigma^{\un{a}})_{\alpha\beta}(\sigma_{\un{a}}^{\ \un{d}\un{e}})^{\gamma\delta}\right] ({\rm D}_{\gamma}{\rm D}_{\delta}\Psi)~~~,  &&\\
R_{\Dot\alpha\Dot\beta}^{\ \ \ \un{d}\un{e}} ~=~ & -i\frac{6}{5}(\sigma^{[\un{d}})_{\Dot\alpha\Dot\beta}
(\partial^{\un{e}]}\Psi) - \frac{1}{40} \left[\frac{1}{3!} (\sigma^{\un{d}\un{e}[3]})_{\Dot\alpha\Dot\beta}(\sigma_{[3]})^{
\Dot\gamma\Dot\delta}+(\sigma^{\un{a}})_{\Dot\alpha\Dot\beta}(\sigma_{\un{a}}^{\ \un{d}\un{e}})^{\Dot\gamma\Dot\delta}\right]({\rm D}_{\Dot\gamma}{\rm D}_{\Dot\delta}\Psi)~~~,  &&\\
R_{\alpha\Dot\beta}^{\ \ \ \un{d}\un{e}} ~=~ & \frac{1}{40} \Big[ - C_{\a\Dot{\b}} (\s^{\un{d}\un{e}})^{\g\Dot{\d}} + (\s^{\un{d}\un{e}})_{\a\Dot{\b}} C^{\g\Dot{\d}} - \frac{1}{2} (\s_{[2]})_{\a\Dot{\b}} (\s^{\un{d}\un{e}[2]})^{\g\Dot{\d}}  \nonumber\\
& \qquad  + \frac{1}{2} (\s^{\un{d}\un{e}[2]})_{\a\Dot{\b}} (\s_{[2]})^{\g\Dot{\d}}  + \frac{1}{4!4!} \epsilon^{\un{d}\un{e}[4][\bar{4}]} (\s_{[4]})_{\a\Dot{\b}} (\s_{[\bar{4}]})^{\g\Dot{\d}}   \Big] ({\rm D}_{\gamma}{\rm D}_{\Dot\delta}\Psi)  ~~~, &&\\
R_{\alpha\un{b}}^{\ \ \ \un{d}\un{e}} ~=~ & -({\rm D}_{\alpha}\partial^{[\un{d}}
\Psi)\delta_{\un{b}}^{\ \un{e}]}+\frac{1}{5}(\sigma^{\un{d}\un{e}})_{\alpha
}^{\ \gamma}(\pa_{\un{b}}{\rm D}_{\gamma}\Psi)  ~~~,  &&\\
R_{\Dot\alpha\un{b}}^{\ \ \ \un{d}\un{e}} ~=~ & -({\rm D}_{\Dot\alpha}\partial^{[\un{d}}
\Psi)\delta_{\un{b}}^{\ \un{e}]}+\frac{1}{5}(\sigma^{\un{d}\un{e}})_{\Dot\alpha
}^{\ \Dot\gamma}(\pa_{\un{b}}{\rm D}_{\Dot\gamma}\Psi)  ~~~,  &&\\
R_{\un{a}\un{b}}^{\ \ \ \un{d}\un{e}} ~=~ & -(\pa_{[\un{a}}\partial^{[\un{d}}
\Psi)\delta_{\un{b}]}^{\ \un{e}]}  ~~~.
\end{align}

\newpage

\section{Linearized Nordstr\" om Supergravity in 10D, $\cal N$ = 2B Supergeometry}

Now for a final time we replicate the discussions as seen in the previous three chapters 
with a beginning of the on-shell description of 10D, $\cal N$ = 2B superspace 
supergravity here.  A set of torsion and curvature supertensors can be written in the form
\be
\eqalign{
{~~~~~}
T_{\a\bar\b}{}^{\un c}~&=~i(\s^{\un c})_{\a\b}~~~,~~~~T_{\a\b}{}^{\un c}~=~
T_{\bar\a\bar\b}{}^{\un c}
~=~0~~~,~~~~T_{\un a\un b}{}^{\un c}~=~0~~~,
{~~~~~~~~~~~~~~~~~~~~~~~~~~~~~~~~~~~}
\cr
T_{\a\b}{}^{\g}~&=~T_{\bar\a\bar\b}{}^{\g}~=~T_{\a\bar\b}{}^{\bar\g}~=~
\Big[\d_{(\a}{}^{\g}\d_{\b)}{}^{\d}~+~(\s^{\un a})_{\a\b}(\s_{\un a})^{\g\d}\Big]\L_{\d}~~~,
~\cr
T_{\a\b}{}^{\bar\g}~&=~T_{\bar\a\b}{}^{\g}~=~T_{\bar\a\bar\b}{}^{\bar\g}~=~
\Big[\d_{(\a}{}^{\g}\d_{\b)}{}^{\d}~+~(\s^{\un a})_{\a\b}(\s_{\un a})^{\g\d}\Big]
\Bar\L_{\d}~~~,}
\ee
\be  \eqalign{
{~~~~~}
T_{\a \un b}{}^{\bar\g}~&=~\frac1{24} (\s_{\un b})_{\a\d}(\s^{[3]})^{\d\g}
{\Bigl [ } e^{-2\F}(1+\Bar W)\Bar G_{[3]}-e^{-2\F}(1-\Bar W)G_{[3]}-
i(\s_{[3]})^{\e\l}(\L_{\e}\L_{\l}-\Bar\L_{\e}\Bar\L_{\l})\Bigr ]  \cr
&{~~~~}+\frac1{96}(\s^{[3]})_{\a\d}(\s_{\un b})^{\d\g}(G_{[3]}+\Bar G_{[3]})~~~,    \cr
T_{\bar\a \un b}{}^{\g}~&=~\frac1{24}(\s_{\un b})_{\a\d}(\s^{[3]})^{\d \g}\Bigl [
e^{-2\F}(1+W)G_{[3]}-e^{-2\F}(1-W)\Bar G_{[3]}+ i(\s_{[3]})^{\e\l}(\L_{\e}
\L_{\l}-\Bar\L_{\e}\Bar\L_{\l})\Bigr ]    \cr
&{~~~~}+\frac1{96}(\s^{[3]})_{\a\d}(\s_{\un b})^{\d\g}(G_{[3]}+\Bar 
G_{[3]}) ~~~,\cr
T_{\a \un b}{}^{\g}~&=~-T_{\bar\a \un b}{}^{\bar\g}~=~\frac14(\s_{\un b})_
{\a\d} (\s^{\un d})^{\d\g}\Bigl [ e^{-2\F}\de_{\un d}(W-\Bar W)+ 
i \frac74 (\s_{\un d})^{\e\l}\L_{\e}\Bar\L_{\l}\Bigr ] 
{~~~~~~~~~~} {~~~~~~~~} \cr
&{~~~~~~~~~~~~~~~~~~~~~}+i \frac1{48}(\s_{[4]})_{\a}
{}^{\g}\Bigl [~ \frac18 (\s_{\un b}{}^{[4]})^{\e\l}\L_{\e}\Bar\L_{\l}-
\frac53 e^{-2\F}{\tilde F}_{\un b}{}^{[4]}~\Bigr ]~~~, }  \ee
\be
\eqalign{
{~~~~~~}
R_{\a\b \un c\un d}~=&~ i \frac1{12}(\s_{\un c\un d}{}^{[3]})_{\a\b}
\Bigl\{ e^{-2\F}(1+\Bar W)
\Bar G_{[3]}-e^{-2\F}(1-\Bar W)G_{[3]}-i(\s_{[3]})^{\e\l}\bigl [\L_{\e}
\L_{\l}-\Bar\L_{\e}\Bar\L_{\l}\bigr ] {~~}
\cr
& ~~~~~~~~~~~~~~~~~~~ - \frac14 (G_{[3]}+ \Bar G_{[3]})\Bigr\}  \cr 
& -i \frac12(\s^{\un e})_{\a\b}
\Bigl \{e^{-2\F}(1+\Bar W) \Bar G_{\un c\un d\un e}~ -~e^{-2\F}(1-\Bar W)G_{\un c\un d\un e}-i(\s_{\un c\un d\un e})^{\e\l}\bigl [\L_{\e}
\L_{\l}-\Bar\L_{\e}\Bar\L_{\l}\bigr ]  \cr
& ~~~~~~~~~~~~~~~~~ + \frac14 (G_{\un c\un d\un e}+
\Bar G_{\un c\un d\un e})\Bigr \}~~~,
{~~~~}
} \ee
\be\eqalign{
{~~~~~}
R_{\bar\a\bar\b \un c\un d}~=&~ i\frac1{12}(\s_{\un c\un d}{}^{[3]})_{\a\b}
\Bigl\{ e^{-2\F}(1+ W)  G_{[3]}-e^{-2\F}(1- W)\Bar G_{[3]} + i (
\s_{[3]})^{\e\l}\bigl [\L_{\e} \L_{\l}-\Bar\L_{\e}\Bar\L_{\l}\bigr ]\cr
& ~~~~~~~~~~~~~~~~~~ -~ \frac14 (G_{[3]}+\Bar G_{[3]})\Bigr\}  \cr 
& - i \frac12 (\s^{\un e})_{\a\b}\Bigl\{e^{-2\F}(1+ W) G_{\un c\un d \un e} ~-~ e^{-2\F}(1- W)\Bar G_{\un c\un d\un e}+i(\s_{\un c\un d\un e})^{\e\l}\bigl [\L_{\e}
\L_{\l}-\Bar\L_{\e}\Bar\L_{\l}\bigr ]\cr
& ~~~~~~~~~~~~~~~~~  + \frac14 (G_{\un c\un d\un e}+
\Bar G_{\un c\un d\un e})\Bigr \}~~.} \ee
as was noted in the portion of the work in \cite{TP2} devoted to type IIB supergravity.
We will end our discussion here.  As the astute reader can note the expressions
are of increasing complication.  But the central message of the expressions in
(7.1) - (7.4) is that the on-shell description of the 10D, $\cal N$ = 2B theory exists 
in perfect analogy with the on-shell description of 4D, $\cal N$ = 1 superspace
given by the equations in (3.12).

Now for the covariant derivative operators linear in the conformal compensator $\Psi$ 
and necessary for a Nordstr\" om theory may be given by
\begin{flalign}   {~~~~~~~~~~}
\nabla_{\alpha} ~= &~{\rm D}_{\alpha} + \frac{1}{2}\Psi {\rm D}_{\alpha}+l_0(\sigma^{
\un{a}\un{b}})_{\alpha}^{\ \beta}({\rm D}_{\beta}\Psi){\cal M}_{\un{a}\un{b}}&&\\
\nabla_{\bar{\alpha}} ~= &~\Bar{\rm D}_{{\alpha}} + \frac{1}{2}\Bar{\Psi} \Bar{\rm D}_{\alpha}+\Bar{l}_0(\sigma ^{\un{a}\un{b}})_{\alpha}^{\ \beta}(\Bar{\rm 
D}_{\beta}\Bar{\Psi}){\cal M}_{\un{a}\un{b}}&&\\
\nabla_{\un{a}} ~=&~ \pa_{\un{a}}+
l_1\Psi\pa_{\un{a}}+
l_2\Bar{\Psi}\pa_{\un{a}}+
il_3(\sigma_{\un{a}})^{\alpha \beta}({\rm D}_{\alpha}\Bar{\Psi})\Bar{\rm D}_{
\beta}+ il_4(\sigma_{\un{a}})^{\alpha \beta}(\Bar{\rm D}_{{\alpha}}
\Psi){\rm D}_{\beta}\nonumber&&\\
&+il_5(\sigma_{\un{a}})^{\alpha \beta}({\rm D}_{\alpha}\Psi)\Bar{\rm D}_{
{\beta}} +il_6(\sigma_{\un{a}})^{\alpha \beta}(\Bar{\rm D}_{{\alpha}}\Bar{
\Psi}){\rm D}_{\beta}\nonumber&&\\
&+il_7(\sigma_{\un{a}}^{\ \un{d}\un{e}})^{\alpha\beta}({\rm D}_{\alpha}\Bar{\rm 
D}_{\beta}\Bar{\Psi}){\cal M}_{\un{d}\un{e}}+ il_8(\sigma_{\un{a}}^{\ \un{
d}\un{e}})^{\alpha\beta}(\Bar{\rm D}_{{\alpha}}{\rm D}_{\beta}\Psi){\cal 
M}_{\un{d}\un{e}}
\nonumber&&\\
&+
l_9(\pa_{\un{c}}\Psi){\cal M}_{\un{a}}^{\ \un{c}}+
l_{10}(\pa_{\un{c}}\Bar{\Psi}){\cal M}_{\un{a}}^{\ \un{c}}
\end{flalign}
with the Type IIB supersymmetry algebra
\begin{align}
\{{\rm D}_{\alpha},{\rm D}_{\beta}\} ~&=~ 0 ~~~,~~~~
\{\Bar{\rm D}_{{\alpha}},\Bar{\rm D}_{{\beta}}\} ~=~ 0 ~~~,~~~~
\{{\rm D}_{\alpha},\Bar{\rm D}_{{\beta}}\} ~=~ i(\sigma^{\un{a}})_{\alpha
\beta}\pa_{\un{a}} ~~~.
\end{align}
 
By adopting the constraints
\begin{align}
T_{\un{a}\un{b}}^{\ \ \un{c}}~=~ 0  ~~~,~~~~ T_{\alpha\bar{\beta}}^{\ 
\ \un{c}} ~=~ i(\sigma^{\un{c}})_{\alpha\beta}  ~~~,
\end{align}
we have the following parameterization results:
$$
l_1 ~=~ l_2 ~=~ \frac{1}{2} ~~,~~
l_3 ~=~ l_4 ~=~ -\frac{1}{32}  ~~,~~
l_5 ~=~  l_6 ~=~ -\frac{27}{160} ~~,~~
$$
\begin{align}
l_7 ~=~  l_8~=~0~~ ~~,~~
l_9 ~=~ l_{10} ~=~ -\frac{1}{2} ~~.~~
\end{align}

As the consequence of this choice of parameters, we find the torsion supertensors given in (7.11)
- (7.28). 
\begin{align} {~~~~~~~~~~}
T_{\a\b}^{\ \ \un{c}} ~=~ & 0 ~~~, &&\\
T_{\a\b}^{\ \ \g} ~=~ & \frac{2}{5}(\sigma^{\un c})_{\a\b}(\sigma_{\un c})^{\g\d}({\rm D}_{\d}\Psi) ~~~, &&\\
T_{\a\b}^{\ \ \bar\g} ~=~ & 0 ~~~, &&\\
T_{\bar\a\bar\b}^{\ \ \un{c}} ~=~ & 0 ~~~, &&\\
T_{\bar\a\bar\b}^{\ \ \g} ~=~ & 0 ~~~, &&\\
T_{\bar\a\bar\b}^{\ \ \bar\g} ~=~ & \frac{2}{5}(\sigma^{\un c})_{\a\b}(\sigma_{\un c})^{\g\d}(\Bar{\rm D}_{\d}\Bar\Psi) ~~~, &&\\
T_{\a\bar\b}^{\ \ \un{c}} ~=~ & i(\sigma^{\un c})_{\a\b} ~~~, &&\\
T_{\a\bar\b}^{\ \ \g} ~=~ \nonumber& -\frac{1}{320}\Big[(\sigma^{[3]})_{\a\b}(\sigma_{[3]})^{\g\d}+\frac{1}{24}(\sigma^{[5]})_{\a\b}(\sigma_{[5]})^{\g\d}\Big](\Bar{\rm D}_{\d}\Bar\Psi)&&\\ 
&+\frac{1}{192}\Big[-(\sigma^{[3]})_{\a\b}(\sigma_{[3]})^{\g\d}+\frac{1}{40}(\sigma^{[5]})_{\a\b}(\sigma_{[5]})^{\g\d}\Big](\Bar{\rm D}_{\d}\Psi) ~~~, &&\\
T_{\a\bar\b}^{\ \ \bar\g} ~=~ \nonumber& -\frac{1}{320}\left[(\sigma^{[3]})_{\a\b}(\sigma_{[3]})^{\g\d}+\frac{1}{24}(\sigma^{[5]})_{\a\b}(\sigma_{[5]})^{\g\d}\right]({\rm D}_{\d}\Psi)&&\\
&+\frac{1}{192}\left[-(\sigma^{[3]})_{\a\b}(\sigma_{[3]})^{\g\d}+\frac{1}{40}(\sigma^{[5]})_{\a\b}(\sigma_{[5]})^{\g\d}\right]({\rm D}_{\d}\Bar\Psi) ~~~, &&\\
T_{\a\un b}^{\ \ \un c}  ~=~ & \left[\frac{53}{160}\d_{\un b}^{\ \un c}\d_{\a}^{\ \g}+\frac{59}{160}(\sigma_{\un b}^{\ \un c})_{\a}^{\ \g}\right]({\rm D}_{\g}\Psi)+\left[\frac{15}{32}\d_{\un b}^{\ \un c}\d_{\a}^{\ \g}+\frac{1}{32}(\sigma_{\un b}^{\ \un c})_{\a}^{\ \g}\right]({\rm D}_{\g}\Bar\Psi) ~~~,  &&\\
T_{\a\un b}^{\ \ \g} ~=~ \nonumber& -\frac{31}{64}\d_{\a}^{\ \g}(\partial_{\un b}\Psi)+\frac{27}{320}\d_{\a}^{\ \g}(\partial_{\un b}\Bar\Psi)+\frac{15}{64}(\sigma_{\un b}^{\ \un c})_{\a}^{\ \g}(\partial_{\un c}\Psi)+\frac{53}{320}(\sigma_{\un b}^{\ \un c})_{\a}^{\ \g}(\partial_{\un c}\Bar\Psi)&&\\ \nonumber
& -i\frac{1}{512}\Big[\frac{1}{2}(\sigma^{[2]})_{\a}^{\ \g}(\sigma_{\un b[2]})^{\b\d} -\frac{1}{3!}(\sigma_{\un b [3]})_{\a}^{\ \g}(\sigma^{[3]})^{\b\d}\Big]({\rm D}_{\b}\Bar{\rm D}_{\d}\Psi)&&\nonumber\\ 
&-i\frac{27}{2560}\Big[\frac{1}{2}(\sigma^{[2]})_{\a}^{\ \g}(\sigma_{\un b[2]})^{\b\d} -\frac{1}{3!}(\sigma_{\un b [3]})_{\a}^{\ \g}(\sigma^{[3]})^{\b\d}\Big]({\rm D}_{\b}\Bar{\rm D}_{\d}\Bar\Psi) ~~~, &&\\
T_{\a\un b}^{\ \ \bar\g} ~=~ \nonumber& -i\frac{1}{512}\left[\frac{1}{2}(\sigma^{[2]})_{\a}^{\ \g}(\sigma_{\un b[2]})^{\b\d}-\frac{1}{3!}(\sigma_{\un b[3]})_{\a}^{\ \g}(\sigma^{[3]})^{\b\d}\right]({\rm D}_{\b}{\rm D}_{\d}\Bar\Psi)&&\\
&-i\frac{27}{2560}\left[\frac{1}{2}(\sigma^{[2]})_{\a}^{\ \g}(\sigma_{\un b[2]})^{\b\d}-\frac{1}{3!}(\sigma_{\un b [3]})_{\a}^{\ \g}(\sigma^{[3]})^{\b\d}\right]({\rm D}_{\b}{\rm D}_{\d}\Psi) ~~~, &&\\
T_{\bar\a\un b}^{\ \ \un c}  ~=~ & \left[\frac{15}{32}\d_{\un b}^{\ \un c}\d_{\a}^{\ \g}+\frac{1}{32}(\sigma_{\un b}^{\ \un c})_{\a}^{\ \g}\right](\Bar{\rm D}_{\g}\Psi)+\left[\frac{53}{160}\d_{\un b}^{\ \un c}\d_{\a}^{\ \g}+\frac{59}{160}(\sigma_{\un b}^{\ \un c})_{\a}^{\ \g}\right](\Bar{\rm D}_{\g}\Bar\Psi) ~~~, &&\\
T_{\bar\a\un b}^{\ \ \g} ~=~ \nonumber& -i\frac{1}{512}\left[\frac{1}{2}(\sigma^{[2]})_{\a}^{\ \g}(\sigma_{\un b [2]})^{\b\d}-\frac{1}{3!}(\sigma_{\un b [3]})_{\a}^{\ \g}(\sigma^{[3]})^{\b\d}\right](\Bar{\rm D}_{\b}\Bar{\rm D}_{\d}\Psi)&&\\
&-i\frac{27}{2560}\left[\frac{1}{2}(\sigma^{[2]})_{\a}^{\ \g}(\sigma_{\un b[2]})^{\b\d}-\frac{1}{3!}(\sigma_{\un b [3]})_{\a}^{\ \g}(\sigma^{[3]})^{\b\d}\right](\Bar{\rm D}_{\b}\Bar{\rm D}_{\d}\Bar\Psi)  ~~~, &&\\
T_{\bar\a\un b}^{\ \ \bar\g} ~=~ \nonumber& -\frac{31}{64}\d_{\a}^{\ \g}(\partial_{\un b}\Bar\Psi)+\frac{27}{320}\d_{\a}^{\ \g}(\partial_{\un b}\Psi)+\frac{15}{64}(\sigma_{\un b}^{\ \un c})_{\a}^{\ \g}(\partial_{\un c}\Bar\Psi)+\frac{53}{320}(\sigma_{\un b}^{\ \un c})_{\a}^{\ \g}(\partial_{\un c}\Psi)&&\\ \nonumber
& -i\frac{1}{512}\Big[\frac{1}{2}(\sigma^{[2]})_{\a}^{\ \g}(\sigma_{\un b[2]})^{\b\d}-\frac{1}{3!}(\sigma_{\un b  [3]})_{\a}^{\ \g}(\sigma^{[3]})^{\b\d}\Big](\Bar{\rm D}_{\b}{\rm D}_{\d}\Bar\Psi)&&\nonumber \\ 
&-i\frac{27}{2560}\Big[\frac{1}{2}(\sigma^{[2]})_{\a}^{\ \g}(\sigma_{\un b[2]})^{\b\d}-\frac{1}{3!}(\sigma_{\un b  [3]})_{\a}^{\ \g}(\sigma^{[3]})^{\b\d}\Big](\Bar{\rm D}_{\b}{\rm D}_{\d}\Psi) ~~~, &&\\
T_{\un{a}\un{b}}^{\ \ \un c} ~=~ & 0 ~~~, &&\\
T_{\un{a}\un{b}}^{\ \ \g} ~=~ & i\frac{1}{32}(\sigma_{[\un{a
}})^{\g\d}(\pa_{\un{b}]}\Bar{\rm D}_{\d}\Psi)+i\frac{27}{160}(\sigma_{[\un{a
}})^{\g\d}(\pa_{\un{b}]}\Bar{\rm D}_{\d}\Bar\Psi) ~~~, &&\\ 
T_{\un{a}\un{b}}^{\ \ \bar\g} ~=~ & i\frac{1}{32}(\sigma_{[\un{a
}})^{\g\d}(\pa_{\un{b}]}{\rm D}_{\d}\Bar\Psi)+i\frac{27}{160}(\sigma_{[\un{a
}})^{\g\d}(\pa_{\un{b}]}{\rm D}_{\d}\Psi) ~~~.
\end{align}

For the components of the curvatures, we find the results seen in (7.29) - (7.34).
\begin{align} {~~~~~~~~~~}
R_{\alpha\beta}^{\ \ \ \un{d}\un{e}} ~=~ & \frac{1}{40}\left[\frac{1}{3!}(\sigma^{\un{d}\un{e}[3]})_{\alpha\beta}(\sigma_{[3]})^{
\gamma\delta}-(\sigma^{\un{a}})_{\alpha\beta}(\sigma_{\un{a}}^{\ \un{d}\un{e}})^{\gamma\delta}\right]({\rm D}_{\gamma}{\rm D}_{\delta}\Psi) ~~~,  &&\\
R_{\bar\alpha\bar\beta}^{\ \ \ \un{d}\un{e}} ~=~ & \frac{1}{40}\left[\frac{1}{3!}(\sigma^{\un{d}\un{e}[3]})_{\a\b}(\sigma_{[3]})^{
\g\d}-(\sigma^{\un{a}})_{\a\b}(\sigma_{\un{a}}^{\ \un{d}\un{e}})^{\g\d}\right](\Bar{\rm D}_{\g}\Bar{\rm D}_{\d}\Bar\Psi)~~~,  &&\\
R_{\alpha\bar\beta}^{\ \ \ \un{d}\un{e}} ~=~ \nonumber& -i\frac{3}{5}(\sigma^{[\un d})_{\a\b}(\partial^{\un e]}(\Psi+\Bar\Psi)) - i\frac{1}{10}(\sigma^{\un{d}\un{e}\un{f}})_{\a\b}(\partial_{\un f}(\Psi+\Bar\Psi)) &&\\ \nonumber
&-\frac{1}{80}\left[(\sigma^{\un{a}})_{\a\b}(\sigma_{\un{a}}^{\ \un{d}\un{e}})^{\g\d} -\frac{1}{2
}(\sigma^{[2][\un{d}})_{\a\b}(\sigma^{\un{e}]}_{\ \ [2]})^{\g\d} -\frac{1}{3!} (\sigma^{\un{d}\un{e}[
3]})_{\a\b}(\sigma_{[3]})^{\g\d}  \right] (\Bar{\rm D}_{\g}{\rm D}_{\d}\Psi) ~~~  &&\\ 
&-\frac{1}{80}\left[(\sigma^{\un{a}})_{\a\b}(\sigma_{\un{a}}^{\ \un{d}\un{e}})^{\g\d} -\frac{1}{2
}(\sigma^{[2][\un{d}})_{\a\b}(\sigma^{\un{e}]}_{\ \ [2]})^{\g\d} -\frac{1}{3!} (\sigma^{\un{d}\un{e}[
3]})_{\a\b}(\sigma_{[3]})^{\g\d}  \right] ({\rm D}_{\g}\Bar{\rm D}_{\d}\Bar\Psi) ~~~,  &&\\
R_{\a\un{b}}^{\ \ \ \un{d}\un{e}} ~=~ & -\frac{1}{2}({\rm D}_{\a}\partial^{[\un{d}}
(\Psi+\Bar{\Psi}))\d_{\un{b}}^{\ \un{e}]} +\frac{1}{5}(\sigma^{\un{d}\un{e}})_{\a
}^{\ \g}(\pa_{\un{b}}{\rm D}_{\g}\Psi)  ~~~,  &&\\
R_{\bar\a\un{b}}^{\ \ \ \un{d}\un{e}} ~=~ & -\frac{1}{2}(\Bar{\rm D}_{\a}\partial^{[\un{d}}
(\Psi+\Bar{\Psi}))\d_{\un{b}}^{\ \un{e}]} +\frac{1}{5}(\sigma^{\un{d}\un{e}})_{\a
}^{\ \g}(\pa_{\un{b}}\Bar{\rm D}_{\g}\Bar\Psi)  ~~~,  &&\\
R_{\un{a}\un{b}}^{\ \ \ \un{d}\un{e}} ~=~ & -\frac{1}{2}(\pa_{[\un{a}}\partial^{[\un{d}}
(\Psi+\Bar{\Psi}))\d_{\un{b}]}^{\ \un{e}]}  ~~~.
\end{align}

\newpage

\section{Conclusion}
\label{conclusions}

 \vskip,2in

This work gives a proposal for descriptions of Nordstr\" om supergravity in eleven and
ten dimensions.  This work is but a foundation as in future extensions of this work
we plan to continue this exploration.  One obvious direction is to extract from this
superfield work the implied component level descriptions that follow from the superfield
equations we have presented.  Our work is based on the assumption that in each of
the cases of 11D, $\cal N$ = 1, 10D, $\cal N$ = 1, 10D, $\cal N$ = 2A and 10D,
$\cal N$ = 2B, a single scalar superfield is required to provide such a description
as this was the case for both ordinary gravitation as well as 4D, $\cal N$ = 1
supergravity.  

Having obtained the results for the theories in ten and eleven dimension superspaces, 
we can compare those results with the ones seen in chapter three.  Looking back
at (\ref{eq:uzw}), with a bit of effort, one can show that the condition $H{}^{\un a}$ =
0 cause only modification in the form of the equations.  Namely the terms $W
{}_{a\ \b \g}$ will vanish under this restriction.  It is thus pointedly seen that all
the basic superfields (i.\ e.\ $R$ and $G{}_{\un a}$) in the algebra of the superspace 
supergravity covariant derivative are bosonic.  This is to be compared to the results
shown in (\ref{eq:uuuu}) where a fermionic superfield $T{}_{\a}$ appears.  In
all of the higher dimensional theories such superfields appear ubiquitously.

In the future we will also address the very important quest of whether
there exists a superspace action for the Nordstr\" om supergravity theories in
high dimension.  It is clear that in order for this to be the case, it is necessary
that the scalar superfield should satisfy some superdifferential constraints.
The expectation is suggested by the structure of the 4D, $\cal N$ = 1 theory.
We remind the reader that the irreducible theories require that the superfield
$X$ is subject to some differential constraints.  So it is natural to expect this
to extend into the higher dimensional theory.

Our approach also raises an interesting question about Superstring Theories,
M-Theory, and F-Theory.  Do these theories also possess consistent truncation
limits
that include Nordstr\" om supergravity theories in their low energy limits?
If the answer is affirmative, such limits might provide laboratories in which
to investigate these more complicated mathematical structures.

 \vspace{.05in}
 \begin{center}
\parbox{4in}{{\it ``The most effective way to do it, is to do it.'' \\ ${~}$ 
\\ ${~}$ }\,\,-\,\, Amelia Earhart $~~~~~~~~~$}
 \parbox{4in}{
 $~~$}  
 \end{center}

  \noindent
{\bf Dedication}\\[.1in] \indent
SJG wishes to dedicate this work to the memory of Shota Ivan Vashakidze,
a valued friend and collaborator in the exploration of ten dimensional
superspace geometry.
 
 \noindent
{\bf Acknowledgements}\\[.1in] \indent
The research of S.\ J.\ Gates, Jr., Y.\ Hu, and S.-N.\ Mak is supported by the endowment 
of the Ford Foundation Professorship of Physics at Brown University and 
partially supported by the U.S. National Science Foundation grant PHY-1315155.

\newpage \noindent
\appendix
\section{11D Clifford Algebra Representation}

In this section we briefly summerize the convention that we adopted for 11D 
gamma matrices. Our 32$\times$ 32 gamma matrices are defined by the 
Clifford algebra:
\be
\{ \gamma^{\un{a}}, \gamma^{\un{b}} \} ~=~ 2 \eta^{\un{a}\un{b}} \, \mathbb{I}~~~,
\ee
where $\mathbb{I}$ denotes the 32$\times$32 identity matrix and
the inverse metric $\eta^{\un{a}\un{b}}$ follows the "most plus" signature:
\be
    \eta^{\un{a}\un{b}} ~=~ \text{diag} ( -1, +1, +1, +1, +1, +1, +1, +1, +1, +1, +1)~~~.
\ee
It is known that D-dimensional space-time Dirac spinor has $2^{\frac{
D-1}{2}}$ components when D is odd, and $2^{D/2}$ components when D is 
even. Hence in 11D, the spinor indices of the gamma matrices, denoted by 
$\alpha$, $\beta$ and so forth, run from 1 to 32.  

One can raise and lower the spinor indices via the "spinor metric", $C_{\alpha
\beta}$, which satisfies:
\be
C_{\alpha\beta}~=~-C_{\beta\alpha} ~~~,~~~
C_{\alpha\beta} C^{\gamma\beta} ~=~ \delta_{\alpha}^{\ \gamma}~~~.
\ee

The gamma matrices with multiple vector indices are defined through the equations:
\begin{align}
\gamma^{\un{a}} \gamma^{\un{b}} =& \gamma^{\un{a}\un{b}} + \eta^{\un{
a}\un{b}}   \\
\gamma^{\un{b}} \gamma^{\un{a}} =& - \gamma^{\un{a}\un{b}} + \eta^{\un{
a}\un{b}}   \\
\gamma^{\un{a}} \gamma^{\un{b}\un{c}} =& \gamma^{\un{a}\un{b}\un{c}} + 
\eta^{\un{a}[\un{b}} \gamma^{\un{c}]}  \\
\gamma^{\un{b}\un{c}} \gamma^{\un{a}} =& \gamma^{\un{a}\un{b}\un{c}} - 
\eta^{\un{a}[\un{b}} \gamma^{\un{c}]}  \\
\gamma^{\un{a}} \gamma^{\un{b}\un{c}\un{d}} =& \gamma^{\un{a}\un{b}
\un{c}\un{d}} + \tfrac{1}{2} \eta^{\un{a}[\un{b}} \gamma^{\un{c}\un{d}]}  \\
\gamma^{\un{b}\un{c}\un{d}} \gamma^{\un{a}} =& - \gamma^{\un{a}\un{b}
\un{c}\un{d}} + \tfrac{1}{2} \eta^{\un{a}[\un{b}} \gamma^{\un{c}\un{d}]}  \\
\gamma^{\un{a}} \gamma^{\un{b}\un{c}\un{d}\un{e}} =& \gamma^{\un{a}
\un{b}\un{c}\un{d}\un{e}} + \tfrac{1}{3!} \eta^{\un{a}[\un{b}} \gamma^{\un{c}
\un{d}\un{e}]}  \\
\gamma^{\un{b}\un{c}\un{d}\un{e}} \gamma^{\un{a}} =& \gamma^{\un{a}
\un{b}\un{c}\un{d}\un{e}} - \tfrac{1}{3!} \eta^{\un{a}[\un{b}} \gamma^{\un{c}
\un{d}\un{e}]}  \\
\gamma^{\un{a}} \gamma^{\un{b}\un{c}\un{d}\un{e}\un{f}} =& \tfrac{1}{5!} 
\epsilon^{\un{a}\un{b}\un{c}\un{d}\un{e}\un{f}[5]} \gamma_{[5]} + \tfrac{1}{
4!} \eta^{\un{a}[\un{b}} \gamma^{\un{c}\un{d}\un{e}\un{f}]}  \\
\gamma^{\un{b}\un{c}\un{d}\un{e}\un{f}} \gamma^{\un{a}} =& - \tfrac{1
}{5!} \epsilon^{\un{a}\un{b}\un{c}\un{d}\un{e}\un{f}[5]} \gamma_{[5]} + 
\tfrac{1}{4!} \eta^{\un{a}[\un{b}} \gamma^{\un{c}\un{d}\un{e}\un{f}]}
\end{align}

The symmetric relations of the gamma matrices are given by:
\begin{align}
(\gamma^{\un{a}})_{\alpha\beta} =& (\gamma^{\un{a}})_{\beta\alpha} \\ 
(\gamma^{\un{a}\un{b}})_{\alpha\beta} =& (\gamma^{\un{a}\un{b}})_{\beta
\alpha} \\ 
(\gamma^{\un{a}\un{b}\un{c}})_{\alpha\beta} =& - (\gamma^{\un{a}\un{b}
\un{c}})_{\beta\alpha} \\ (\gamma^{\un{a}\un{b}\un{c}\un{d}})_{\alpha\beta} 
=& - (\gamma^{\un{a}\un{b}\un{c}\un{d}})_{\beta\alpha} \\
(\gamma^{\un{a}\un{b}\un{c}\un{d}\un{e}})_{\alpha\beta} =& (\gamma^{
\un{a}\un{b}\un{c}\un{d}\un{e}})_{\beta\alpha}
\end{align}

From the definitions, one can easily work out the following trace identities:
\begin{align}
&(\gamma_{\un{a}})_{\alpha}{}^{\beta}(\gamma^{\un{b}})_{\beta}{}^{\alpha} = 
32\delta_{\un{a}}^{\ \un{b}}\\
&(\gamma_{\un{a}\un{b}})_{\alpha}{}^{\beta}(\gamma^{\un{c}\un{d}})_{\beta}{
}^{\alpha} = -32\delta_{[\un{a}}^{\ \un{c}}\delta_{\un{b}]}^{\ \un{d}}\\
&(\gamma_{\un{a}\un{b}\un{c}})_{\alpha}{}^{\beta}(\gamma^{\un{d}\un{e}\un{
f}})_{\beta}{}^{\alpha} = -32\delta_{\un{[a}}^{\ \un{d}}\delta_{\un{b}}^{\ \un{e}}
\delta_{\un{c}]}^{\ \un{f}}\\
&(\gamma_{\un{a}\un{b}\un{c}\un{d}})_{\alpha}{}^{\beta}(\gamma^{\un{e}\un{
f}\un{g}\un{h}})_{\beta}{}^{\alpha} = 32\delta_{[\un{a}}^{\ \un{e}}\delta_{\un{b}
}^{\ \un{f}}\delta_{\un{c}}^{\ \un{g}}\delta_{\un{d}]}^{\ \un{h}}\\
&(\gamma_{\un{a}\un{b}\un{c}\un{d}\un{e}})_{\alpha}{}^{\beta}(\gamma^{
\un{f}\un{g}\un{h}\un{i}\un{j}})_{\beta}{}^{\alpha} = 32 \delta_{\un{[a}}^{\ \un{f}}
\delta_{\un{b}}^{\ \un{g}}\delta_{\un{c}}^{\ \un{h}}\delta_{\un{d}}^{\ \un{i}}
\delta_{\un{e}]}^{\ \un{j}}
\end{align}
as well as the following Fierz identities: 
\begin{align}
    \d_{(\a}^{\ \d}\d_{\b)}^{\ \g} ~=~& \frac{1}{16} \Big\{ - (\g^{\un{c}})_{\a\b} (\g_{\un{c}})^{\d\g} + \frac{1}{2} (\g^{[2]})_{\a\b} (\g_{[2]})^{\d\g} - \frac{1}{5!} (\g^{[5]})_{\a\b} (\g_{[5]})^{\d\g} \Big\}   \\ 
    (\g^{[2]})_{(\a}^{\ \d} (\g_{[2]})_{\b)}^{\ \g} ~=~& \frac{1}{16} \Big\{ - 70(\g^{\un{c}})_{\a\b} (\g_{\un{c}})^{\d\g} + 19 (\g^{[2]})_{\a\b} (\g_{[2]})^{\d\g} - \frac{1}{12} (\g^{[5]})_{\a\b} (\g_{[5]})^{\d\g} \Big\}  \\
\begin{split}
    \d_{\a}^{\ [\d} (\g_{\un{b}})^{\e]\g} ~=~& \frac{1}{16} \Big\{ - (\g_{\un{b}})_{\a}^{\ \g} C^{\d\e} + \frac{1}{2} (\g^{[2]})_{\a}^{\ \g} (\g_{\un{b}[2]})^{\d\e} - \frac{1}{3!} (\g_{\un{b}[3]})_{\a}^{\ \g} (\g^{[3]})^{\d\e}  \\
    & \qquad  + \frac{1}{3!} (\g^{[3]})_{\a}^{\ \g} (\g_{\un{b}[3]})^{\d\e} - \frac{1}{4!} (\g_{\un{b}[4]})_{\a}^{\ \g} (\g^{[4]})^{\d\e}   \Big\}
\end{split} \\
\begin{split}
    (\g^{\un{d}\un{e}})_{(\a}^{\ \ \g}\d_{\b)}^{\ \ \d} ~=~& \frac{1}{16}  \Big\{  (\g_{[1]})_{\a\b} (\g^{[1]\un{d}\un{e}})^{\g\d} - (\g^{[\un{d}})_{\a\b} (\g^{\un{e}]})^{\g\d} \\
    & \qquad  - \frac{1}{2} (\g_{[2]})_{\a\b} (\g^{[2]\un{d}\un{e}})^{\g\d} - (\g^{[1] [\un{d}})_{\a\b} (\g^{\un{e}]}{}_{[1]})^{\g\d} + (\g^{\un{d}\un{e}})_{\a\b} C^{\g\d}  \\
    & \qquad  + \frac{1}{5!4!} \e^{\un{d}\un{e}[5][4]} (\g_{[5]})_{\a\b} (\g_{[4]})^{\g\d}  - \frac{1}{4!} (\g^{[4] [\un{d}})_{\a\b} (\g^{\un{e}]}{}_{[4]})^{\g\d} - \frac{1}{3!} (\g^{\un{d}\un{e}[3]})_{\a\b} (\g_{[3]})^{\g\d}   \Big\}
\end{split}
\end{align}
Finally, we list the explicit representations of 11D gamma matrices in terms of 
tensor products of Pauli matrices:

Spinor metric: 
\begin{equation}
 C_{\alpha\beta} = - i \sigma^{2} \otimes \mathbb{I}_{2} \otimes \mathbb{I
 }_{2} \otimes \mathbb{I}_{2} \otimes \mathbb{I}_{2}
\end{equation}
Gamma matrices:
\begin{align}
(\gamma^{\un{0}})_{\alpha}{}^{\beta} =& i \sigma^{2} \otimes \mathbb{I}_{
2} \otimes \mathbb{I}_{2} \otimes \mathbb{I}_{2} \otimes \mathbb{I}_{2}   \\
(\gamma^{\un{1}})_{\alpha}{}^{\beta} =& \sigma^{1} \otimes \sigma^{2} 
\otimes \sigma^{2} \otimes \sigma^{2} \otimes \sigma^{2}   \\
(\gamma^{\un{2}})_{\alpha}{}^{\beta} =& \sigma^{1} \otimes \sigma^{2} 
\otimes \sigma^{2} \otimes \mathbb{I}_{2} \otimes \sigma^{1}   \\
(\gamma^{\un{3}})_{\alpha}{}^{\beta} =& \sigma^{1} \otimes \sigma^{2} 
\otimes \sigma^{2} \otimes \mathbb{I}_{2} \otimes \sigma^{3}   \\
(\gamma^{\un{4}})_{\alpha}{}^{\beta} =& \sigma^{1} \otimes \sigma^{2} 
\otimes \sigma^{1} \otimes \sigma^{2} \otimes \mathbb{I}_{2}   \\
(\gamma^{\un{5}})_{\alpha}{}^{\beta} =& \sigma^{1} \otimes \sigma^{2} 
\otimes \sigma^{3} \otimes \sigma^{2} \otimes \mathbb{I}_{2}   \\
(\gamma^{\un{6}})_{\alpha}{}^{\beta} =& \sigma^{1} \otimes \sigma^{2} 
\otimes \mathbb{I}_{2} \otimes \sigma^{1} \otimes \sigma^{2}   \\
(\gamma^{\un{7}})_{\alpha}{}^{\beta} =& \sigma^{1} \otimes \sigma^{2} 
\otimes \mathbb{I}_{2} \otimes \sigma^{3} \otimes \sigma^{2}   \\
(\gamma^{\un{8}})_{\alpha}{}^{\beta} =& \sigma^{1} \otimes \sigma^{1} 
\otimes \mathbb{I}_{2} \otimes \mathbb{I}_{2} \otimes \mathbb{I}_{2}   \\
(\gamma^{\un{9}})_{\alpha}{}^{\beta} =& \sigma^{1} \otimes \sigma^{3} 
\otimes \mathbb{I}_{2} \otimes \mathbb{I}_{2} \otimes \mathbb{I}_{2}   \\
(\gamma^{\un{10}})_{\alpha}{}^{\beta} =& \sigma^{3} \otimes \mathbb{
I}_{2} \otimes \mathbb{I}_{2} \otimes \mathbb{I}_{2} \otimes \mathbb{I}_{2}
\end{align}

\newpage
\newpage \noindent
\section{10D Clifford Algebra Representation}

In this section we briefly summerize the convention that we adopted for 
10D sigma matrices. The Clifford algebra is
\begin{equation}
(\sigma^{\un{a}})_{\alpha\beta} (\sigma^{\un{b}})^{\beta\gamma} + (\sigma^{
\un{b}})_{\alpha\beta} (\sigma^{\un{a}})^{\beta\gamma} = 2 \eta^{\un{a}\un{
b}} \delta_{\alpha}{}^{\gamma}~~~,
\end{equation}
where the inverse metric $\eta^{\un{a}\un{b}}$ is:
\begin{equation}
\eta^{\un{a}\un{b}} = \text{diag} ( -1, +1, +1, +1, +1, +1, +1, +1, +1, +1)~~~.
\end{equation}
In 10D, the Dirac spinor has $2^{D/2} = 32$ components. We use undotted 
Greek index to denote 16 component left-handed Majorana spinor, and 
dotted index to denote right-handed ones, 
\begin{equation}
(\psi^{\alpha})^{*} = \psi^{\alpha}~~~,  \qquad  (\psi^{\Dot{\alpha}})^{*} = \psi^{
\Dot{\alpha}}
\end{equation}
where $\alpha = 1, \ldots, 16$ and $\Dot{\alpha} = 1, \ldots, 16$. We raise 
and lower the spinor indices by spinor metric $C_{\alpha\Dot{\beta}}$ as 
follows:
\begin{equation}
\begin{aligned}
& \psi_{\Dot{\beta}} = \psi^{\alpha} C_{\alpha\Dot{\beta}}~~~,  & & \psi_{\alpha} 
= \psi^{\Dot{\beta}} C_{\alpha\Dot{\beta}}~~~, \\
&  C_{\alpha\Dot{\gamma}} C^{\alpha\Dot{\beta}} = \delta_{\Dot{\gamma}}{
}^{\Dot{\beta}}~~~,  & & C_{\gamma\Dot{\beta}} C^{\alpha\Dot{\beta}} = \delta_{
\gamma}{}^{\alpha}~~~.
\end{aligned}
\end{equation}
The sigma matrices are bispinors. There are three types of them: purely 
left-handed:
\begin{equation}
(\sigma^{\un{a}})_{\alpha\beta}~~~,  \qquad  (\sigma^{\un{a}\un{b}\un{c}})_{
\alpha\beta}~~~, \qquad  (\sigma^{\un{a}\un{b}\un{c}\un{d}\un{e}})_{\alpha
\beta}~~~;
\end{equation}
purely right-handed (related to purely left-handed by the following):
\begin{equation}
(\sigma^{\un{a}})^{\alpha\beta} = C^{\alpha\Dot{\alpha}} C^{\beta\Dot{\beta
}}(\sigma^{\un{a}})_{\Dot{\alpha}\Dot{\beta}}~~~,  \qquad  (\sigma^{\un{a}\un{b}
\un{c}})^{\alpha\beta} = C^{\alpha\Dot{\alpha}} C^{\beta\Dot{\beta}} (\sigma^{
\un{a}\un{b}\un{c}})_{\Dot{\alpha}\Dot{\beta}}~~~,  \qquad  (\sigma^{\un{a}\un{b}
\un{c}\un{d}\un{e}})^{\alpha\beta} = C^{\alpha\Dot{\alpha}} C^{\beta\Dot{
\beta}} (\sigma^{\un{a}\un{b}\un{c}\un{d}\un{e}})_{\Dot{\alpha}\Dot{\beta}}~~~;
\end{equation}
and mixed bispinors:
\begin{equation}
C_{\alpha\Dot{\beta}}~~~,  \qquad  (\sigma^{\un{a}\un{b}})_{\alpha\Dot{\beta
}}~~~,  \qquad  (\sigma^{\un{a}\un{b}\un{c}\un{d}})_{\alpha\Dot{\beta}}~~~,
\end{equation}
which have relations 
\begin{equation}
\delta_{\alpha}{}^{\beta} = C^{\beta\Dot{\beta}} C_{\alpha\Dot{\beta}}~~~,  \qquad  
(\sigma^{\un{a}\un{b}})_{\alpha}{}^{\beta} = C^{\beta\Dot{\beta}} (\sigma^{\un{
a}\un{b}})_{\alpha\Dot{\beta}}~~~,  \qquad  (\sigma^{\un{a}\un{b}\un{c}\un{d}})_{
\alpha}{}^{\beta} = C^{\beta\Dot{\beta}} (\sigma^{\un{a}\un{b}\un{c}\un{d}})_{
\alpha\Dot{\beta}}~~~.
\end{equation}
Definition of $\sigma$-matrices with more Lorentz indices:
\begin{align}
(\sigma^{\un{a}})_{\alpha\beta} (\sigma^{\un{b}})^{\beta\gamma} =& (\sigma^{
\un{a}\un{b}})_{\alpha}{}^{\gamma} + \eta^{\un{a}\un{b}} \delta_{\alpha}{}^{
\gamma}  \\
(\sigma^{\un{b}})_{\alpha\beta} (\sigma^{\un{a}})^{\beta\gamma} =& - (\sigma^{
\un{a}\un{b}})_{\alpha}{}^{\gamma} + \eta^{\un{a}\un{b}} \delta_{\alpha}{}^{\gamma}  
\\
(\sigma^{\un{a}})^{\alpha\beta} (\sigma^{\un{b}\un{c}})_{\beta}{}^{\gamma} =& 
(\sigma^{\un{a}\un{b}\un{c}})^{\alpha\gamma} + \eta^{\un{a}[\un{b}} (\sigma^{\un{
c}]})^{\alpha\gamma}  \\
(\sigma^{\un{b}\un{c}})_{\alpha}{}^{\beta} (\sigma^{\un{a}})_{\beta\gamma} =& 
(\sigma^{\un{a}\un{b}\un{c}})_{\alpha\gamma} - \eta^{\un{a}[\un{b}} (\sigma^{\un{
c}]})_{\alpha\gamma}  \\
(\sigma^{\un{a}})_{\alpha\beta} (\sigma^{\un{b}\un{c}\un{d}})^{\beta\gamma} =& 
(\sigma^{\un{a}\un{b}\un{c}\un{d}})_{\alpha}{}^{\gamma} + \tfrac{1}{2} \eta^{\un{
a}[\un{b}} (\sigma^{\un{c}\un{d}]})_{\alpha}{}^{\gamma}  \\
(\sigma^{\un{b}\un{c}\un{d}})_{\alpha\beta} (\sigma^{\un{a}})^{\beta\gamma} =& 
- (\sigma^{\un{a}\un{b}\un{c}\un{d}})_{\alpha}{}^{\gamma} + \tfrac{1}{2} \eta^{\un{
a}[\un{b}} (\sigma^{\un{c}\un{d}]})_{\alpha}{}^{\gamma}  \\
(\sigma^{\un{a}})^{\alpha\beta} (\sigma^{\un{b}\un{c}\un{d}\un{e}})_{\beta}{}^{
\gamma} =& (\sigma^{\un{a}\un{b}\un{c}\un{d}\un{e}})^{\alpha\gamma} + \tfrac{
1}{3!} \eta^{\un{a}[\un{b}} (\sigma^{\un{c}\un{d}\un{e}]})^{\alpha\gamma}  \\
(\sigma^{\un{b}\un{c}\un{d}\un{e}})_{\alpha}{}^{\beta} (\sigma^{\un{a}})_{\beta
\gamma} =& (\sigma^{\un{a}\un{b}\un{c}\un{d}\un{e}})_{\alpha\gamma} - \tfrac{
1}{3!} \eta^{\un{a}[\un{b}} (\sigma^{\un{c}\un{d}\un{e}]})_{\alpha\gamma}  \\
(\sigma^{\un{a}})_{\alpha\beta} (\sigma^{\un{b}\un{c}\un{d}\un{e}\un{f}})^{\beta
\gamma} =& \tfrac{1}{4!} \epsilon^{\un{a}\un{b}\un{c}\un{d}\un{e}\un{f}[4]} (
\sigma_{[4]})_{\alpha}{}^{\gamma} + \tfrac{1}{4!} \eta^{\un{a}[\un{b}} (\sigma^{
\un{c}\un{d}\un{e}\un{f}]})_{\alpha}{}^{\gamma}  \\
(\sigma^{\un{b}\un{c}\un{d}\un{e}\un{f}})_{\alpha\beta} (\sigma^{\un{a}})^{\beta
\gamma} =& - \tfrac{1}{4!} \epsilon^{\un{a}\un{b}\un{c}\un{d}\un{e}\un{f}[4]} (
\sigma_{[4]})_{\alpha}{}^{\gamma} + \tfrac{1}{4!} \eta^{\un{a}[\un{b}} (\sigma^{
\un{c}\un{d}\un{e}\un{f}]})_{\alpha}{}^{\gamma}
\end{align}
and
\begin{align}
(\sigma^{\un{a}})_{\Dot{\alpha}\Dot{\beta}} (\sigma^{\un{b}})^{\Dot{\beta}\Dot{
\gamma}} =& (\sigma^{\un{a}\un{b}})_{\Dot{\alpha}}{}^{\Dot{\gamma}} + \eta^{
\un{a}\un{b}} \delta_{\Dot{\alpha}}{}^{\Dot{\gamma}}  \\
(\sigma^{\un{b}})_{\Dot{\alpha}\Dot{\beta}} (\sigma^{\un{a}})^{\Dot{\beta}\Dot{
\gamma}} =& - (\sigma^{\un{a}\un{b}})_{\Dot{\alpha}}{}^{\Dot{\gamma}} + 
\eta^{\un{a}\un{b}} \delta_{\Dot{\alpha}}{}^{\Dot{\gamma}}  \\
(\sigma^{\un{a}})^{\Dot{\alpha}\Dot{\beta}} (\sigma^{\un{b}\un{c}})_{\Dot{\beta
}}{}^{\Dot{\gamma}} =& (\sigma^{\un{a}\un{b}\un{c}})^{\Dot{\alpha}\Dot{\gamma
}} + \eta^{\un{a}[\un{b}} (\sigma^{\un{c}]})^{\Dot{\alpha}\Dot{\gamma}}  \\
(\sigma^{\un{b}\un{c}})_{\Dot{\alpha}}{}^{\Dot{\beta}} (\sigma^{\un{a}})_{\Dot{
\beta}\Dot{\gamma}} =& (\sigma^{\un{a}\un{b}\un{c}})_{\Dot{\alpha}\Dot{
\gamma}} - \eta^{\un{a}[\un{b}} (\sigma^{\un{c}]})_{\Dot{\alpha}\Dot{\gamma
}}  \\
(\sigma^{\un{a}})_{\Dot{\alpha}\Dot{\beta}} (\sigma^{\un{b}\un{c}\un{d}})^{\Dot
{\beta}\Dot{\gamma}} =& (\sigma^{\un{a}\un{b}\un{c}\un{d}})_{\Dot{\alpha}}{
}^{\Dot{\gamma}} + \tfrac{1}{2} \eta^{\un{a}[\un{b}} (\sigma^{\un{c}\un{d}]})_{
\Dot{\alpha}}{}^{\Dot{\gamma}}  \\
(\sigma^{\un{b}\un{c}\un{d}})_{\Dot{\alpha}\Dot{\beta}} (\sigma^{\un{a}})^{\Dot
{\beta}\Dot{\gamma}} =& - (\sigma^{\un{a}\un{b}\un{c}\un{d}})_{\Dot{\alpha}}{
}^{\Dot{\gamma}} + \tfrac{1}{2} \eta^{\un{a}[\un{b}} (\sigma^{\un{c}\un{d}]})_{
\Dot{\alpha}}{}^{\Dot{\gamma}}  \\
(\sigma^{\un{a}})^{\Dot{\alpha}\Dot{\beta}} (\sigma^{\un{b}\un{c}\un{d}\un{e}}
)_{\Dot{\beta}}{}^{\Dot{\gamma}} =& (\sigma^{\un{a}\un{b}\un{c}\un{d}\un{e}}
)^{\Dot{\alpha}\Dot{\gamma}} + \tfrac{1}{3!} \eta^{\un{a}[\un{b}} (\sigma^{\un{
c}\un{d}\un{e}]})^{\Dot{\alpha}\Dot{\gamma}}  \\
(\sigma^{\un{b}\un{c}\un{d}\un{e}})_{\Dot{\alpha}}{}^{\Dot{\beta}} (\sigma^{\un{
a}})_{\Dot{\beta}\Dot{\gamma}} =& (\sigma^{\un{a}\un{b}\un{c}\un{d}\un{e}})_{
\Dot{\alpha}\Dot{\gamma}} - \tfrac{1}{3!} \eta^{\un{a}[\un{b}} (\sigma^{\un{c}\un{
d}\un{e}]})_{\Dot{\alpha}\Dot{\gamma}}  \\
(\sigma^{\un{a}})_{\Dot{\alpha}\Dot{\beta}} (\sigma^{\un{b}\un{c}\un{d}\un{e
}\un{f}})^{\Dot{\beta}\Dot{\gamma}} =& - \tfrac{1}{4!} \epsilon^{\un{a}\un{b}\un{
c}\un{d}\un{e}\un{f}[4]} (\sigma_{[4]})_{\Dot{\alpha}}{}^{\Dot{\gamma}} + \tfrac{
1}{4!} \eta^{\un{a}[\un{b}} (\sigma^{\un{c}\un{d}\un{e}\un{f}]})_{\Dot{\alpha}}{
}^{\Dot{\gamma}}  \\
(\sigma^{\un{b}\un{c}\un{d}\un{e}\un{f}})_{\Dot{\alpha}\Dot{\beta}} (\sigma^{
\un{a}})^{\Dot{\beta}\Dot{\gamma}} =&  \tfrac{1}{4!} \epsilon^{\un{a}\un{b}\un{
c}\un{d}\un{e}\un{f}[4]} (\sigma_{[4]})_{\Dot{\alpha}}{}^{\Dot{\gamma}} + \tfrac{
1}{4!} \eta^{\un{a}[\un{b}} (\sigma^{\un{c}\un{d}\un{e}\un{f}]})_{\Dot{\alpha}
}{}^{\Dot{\gamma}} 
\end{align}
The sigma matrices with five vector indices satisfy the self-dual / anti-self-dual 
identities:
\begin{align}
(\sigma_{[5]})_{\alpha\beta} =& \frac{1}{5!}\epsilon_{[5]}^{\ \ [\bar{5}]}(\sigma_{
[\bar{5}]})_{\alpha\beta} \\
(\sigma_{[5]})^{\alpha\beta} =& -\frac{1}{5!}\epsilon_{[5]}^{\ \ [\bar{5}]}(\sigma_{
[\bar{5}]})^{\alpha\beta} \\
(\sigma_{[5]})_{\Dot{\alpha}\Dot{\beta}} =& -\frac{1}{5!}\epsilon_{[5]}^{\ \ [\bar{5}
]}(\sigma_{[\bar{5}]})_{\Dot{\alpha}\Dot{\beta}} \\
(\sigma_{[5]})^{\Dot{\alpha}\Dot{\beta}} =& \frac{1}{5!}\epsilon_{[5]}^{\ \ [\bar{
5}]}(\sigma_{[\bar{5}]})^{\Dot{\alpha}\Dot{\beta}}
\end{align}
The symmetric relations of the gamma matrices are given by:
\begin{align}
(\sigma^{\un{a}})_{\alpha\beta} =& (\sigma^{\un{a}})_{\beta\alpha} \\ 
(\sigma^{\un{a}\un{b}\un{c}})_{\alpha\beta} =& - (\sigma^{\un{a}\un{b
}\un{c}})_{\beta\alpha} \\ 
(\sigma^{\un{a}\un{b}\un{c}\un{d}\un{e}})_{\alpha\beta} =& (\sigma^{
\un{a}\un{b}\un{c}\un{d}\un{e}})_{\beta\alpha}
\end{align}
From the definition, we can easily work out the trace identities:
\begin{align}
&(\sigma_{\un{a}})_{\alpha\beta}(\sigma^{\un{b}})^{\alpha\beta} = 16\delta_{
\un{a}}^{\ \un{b}},\\
&(\sigma_{\un{a}\un{b}})_{\alpha}^{\ \beta}(\sigma^{\un{c}\un{d}})_{\beta}^{\ 
\alpha} = -16\delta_{[\un{a}}^{\ \un{c}}\delta_{\un{b}]}^{\ \un{d}},\\
&(\sigma_{\un{a}\un{b}\un{c}})_{\alpha\beta}(\sigma^{\un{d}\un{e}\un{f}})^{
\alpha\beta} = 16\delta_{\un{[a}}^{\ \un{d}}\delta_{\un{b}}^{\ \un{e}}\delta_{
\un{c}]}^{\ \un{f}},\\
&(\sigma_{\un{a}\un{b}\un{c}\un{d}})_{\alpha}^{\ \beta}(\sigma^{\un{e}\un{
f}\un{g}\un{h}})_{\beta}^{\ \alpha} = 16\delta_{[\un{a}}^{\ \un{e}}\delta_{\un
{b}}^{\ \un{f}}\delta_{\un{c}}^{\ \un{g}}\delta_{\un{d}]}^{\ \un{h}},\\
&(\sigma_{\un{a}\un{b}\un{c}\un{d}\un{e}})_{\alpha\beta}(\sigma^{\un{f}\un{
g}\un{h}\un{i}\un{j}})^{\alpha\beta} = 16[\delta_{\un{[a}}^{\ \un{f}}\delta_{\un{
b}}^{\ \un{g}}\delta_{\un{c}}^{\ \un{h}}\delta_{\un{d}}^{\ \un{i}}\delta_{\un{e}]
}^{\ \un{j}}+\epsilon_{\un{a}\un{b}\un{c}\un{d}\un{e}}^{\ \ \ \ \ \ \un{f}\un{g}\un{
h}\un{i}\un{j}}],
\end{align}
and
\begin{align}
&(\sigma_{\un{a}})_{\Dot{\alpha}\Dot{\beta}}(\sigma^{\un{b}})^{\Dot{\alpha}\Dot{
\beta}} = 16\delta_{\un{a}}^{\ \un{b}},\\
&(\sigma_{\un{a}\un{b}})_{\Dot{\alpha}}^{\ \Dot{\beta}}(\sigma^{\un{c}\un{d}})_{
\Dot{\beta}}^{\ \Dot{\alpha}} = -16\delta_{[\un{a}}^{\ \un{c}}\delta_{\un{b}]}^{\ \un{
d}},\\
&(\sigma_{\un{a}\un{b}\un{c}})_{\Dot{\alpha}\Dot{\beta}}(\sigma^{\un{d}\un{e
}\un{f}})^{\Dot{\alpha}\Dot{\beta}} = 16\delta_{\un{[a}}^{\ \un{d}}\delta_{\un{b}
}^{\ \un{e}}\delta_{\un{c}]}^{\ \un{f}},\\
&(\sigma_{\un{a}\un{b}\un{c}\un{d}})_{\Dot{\alpha}}^{\ \Dot{\beta}}(\sigma^{
\un{e}\un{f}\un{g}\un{h}})_{\Dot{\beta}}^{\ \Dot{\alpha}} = 16\delta_{[\un{a}}^{\ 
\un{e}}\delta_{\un{b}}^{\ \un{f}}\delta_{\un{c}}^{\ \un{g}}\delta_{\un{d}]}^{\ \un{
h}},\\
&(\sigma_{\un{a}\un{b}\un{c}\un{d}\un{e}})_{\Dot{\alpha}\Dot{\beta}}(\sigma^{
\un{f}\un{g}\un{h}\un{i}\un{j}})^{\Dot{\alpha}\Dot{\beta}} = 16[\delta_{\un{[a}}^{\ 
\un{f}}\delta_{\un{b}}^{\ \un{g}}\delta_{\un{c}}^{\ \un{h}}\delta_{\un{d}}^{\ \un{i
}}\delta_{\un{e}]}^{\ \un{j}}-\epsilon_{\un{a}\un{b}\un{c}\un{d}\un{e}}^{\ \ \ \ \ \ 
\un{f}\un{g}\un{h}\un{i}\un{j}}].
\end{align}

From the definition, we can also derive the following 10D sigma matrices 
identities:
\begin{align}
(\sigma_{\un{a}})^{\delta\gamma} (\sigma^{[2]})_{\gamma}^{\ \alpha} \sigma_{
[2]})_{\delta}^{\ \beta} =& 54(\sigma_{\un{a}})^{\alpha\beta} \\
(\sigma_{\un{a}})^{\delta\gamma}(\sigma_{\un{c}}^{\ \un{d}})_{\delta}^{\ 
\alpha} (\sigma^{\un{c}\un{e}})_{\gamma}^{\ \beta} =& 6(\sigma_{\un{a}}^{\ 
\un{d}\un{e}})^{\alpha\beta} + 7\eta^{\un{d}\un{e}}(\sigma_{\un{a}})^{\alpha
\beta} - 8 \delta_{\un{a}}^{\ (\un{d}} (\sigma^{\un{e})})^{\alpha\beta} \\
(\sigma_{[5]})^{\alpha\beta}(\sigma^{[2]})_{\alpha}^{\ \delta}(\sigma_{[2]}
)_{\beta}^{\ \gamma} =& - 10 (\sigma_{[5]})^{\delta\gamma} \\
(\sigma_{[5]})^{\Dot{\alpha}\Dot{\beta}} (\sigma^{[2]})_{\Dot{\alpha}}^{\ 
\Dot{\delta}} (\sigma_{[2]})_{\Dot{\beta}}^{\ \Dot{\gamma}} =& - 10 (
\sigma_{[5]})^{\Dot{\delta}\Dot{\gamma}} \\
(\sigma_{[5]})_{\alpha\beta} (\sigma^{\un{c}})^{\alpha\delta}(\sigma_{
\un{c}})^{\beta\gamma} =& 0
\end{align}

as well as the  following Fierz identities:
\begingroup
\allowdisplaybreaks
\begin{align}
    \d_{\a}^{\ \d}\d_{\b}^{\ \g} ~=~& \frac{1}{16} \Big\{ (\s^{\un{c}})_{\a\b} (\s_{\un{c}})^{\d\g} + \frac{1}{3!} (\s^{[3]})_{\a\b} (\s_{[3]})^{\d\g}  +  \frac{1}{2\times 5!} (\s^{[5]})_{\a\b} (\s_{[5]})^{\d\g} \Big\}  \\
    (\s^{[2]})_{\a}^{\ \d} (\s_{[2]})_{\b}^{\ \g} ~=~& \frac{27}{8} (\s^{\un{c}})_{\a\b} (\s_{\un{c}})^{\d\g} + \frac{1}{16} (\s^{[3]})_{\a\b} (\s_{[3]})^{\d\g} - \frac{1}{384} (\s^{[5]})_{\a\b} (\s_{[5]})^{\d\g} \\ 
\begin{split}
    \d_{\a}^{\ \b} (\s_{\un{b}})^{\g\d} ~=~& \frac{1}{16} \Big\{ \d_{\a}^{\ \g} (\s_{\un{b}})^{\b\d} + \frac{1}{2} (\s^{[2]})_{\a}^{\ \g} (\s_{\un{b}[2]})^{\b\d} - (\s_{\un{b}[1]})_{\a}^{\ \g} (\s^{[1]})^{\b\d} \\
    & \qquad  + \frac{1}{4!} (\s^{[4]})_{\a}^{\ \g} (\s_{\un{b}[4]})^{\b\d} - \frac{1}{3!} (\s_{\un{b}[3]})_{\a}^{\ \g} (\s^{[3]})^{\b\d}  \Big\}
\end{split} \\
\begin{split}
    (\s^{\un{d}\un{e}})_{\a}^{\ \g}\d_{\b}^{\ \d} ~=~& \frac{1}{16}  \Big\{  - (\s_{[1]})_{\a\b} (\s^{[1]\un{d}\un{e}})^{\g\d} +(\s^{[\un{d}})_{\a\b} (\s^{\un{e}]})^{\g\d} \\
    & \qquad  - \frac{1}{3!} (\s_{[3]})_{\a\b} (\s^{[3]\un{d}\un{e}})^{\g\d} + \frac{1}{2} (\s^{[2][\un{d}})_{\a\b} (\s^{\un{e}]}_{\ \ [2]})^{\g\d}  + (\s^{\un{d}\un{e}[1]})_{\a\b} (\s_{[1]})^{\g\d} \\
    & \qquad  + \frac{1}{2\times 4!} (\s_{[4]}^{\ \ [\un{d}})_{\a\b} (\s^{\un{e}][4]})^{\g\d} + \frac{1}{3!} (\s^{\un{d}\un{e}[3]})_{\a\b} (\s_{[3]})^{\g\d}   \Big\}
\end{split}  \\
\begin{split}
    (\s^{\un{d}\un{e}})_{(\a}^{\ \ \g}\d_{\b)}^{\ \ \d} ~=~& \frac{1}{16}  \Big\{  - 2 (\s_{[1]})_{\a\b} (\s^{[1]\un{d}\un{e}})^{\g\d} + 2 (\s^{[\un{d}})_{\a\b} (\s^{\un{e}]})^{\g\d} \\
    & \qquad  + \frac{1}{4!} (\s_{[4]}^{\ \ [\un{d}})_{\a\b} (\s^{\un{e}][4]})^{\g\d} + \frac{1}{3} (\s^{\un{d}\un{e}[3]})_{\a\b} (\s_{[3]})^{\g\d}   \Big\}
\end{split}  \\
\begin{split}
    (\s_{\un{a}\un{b}})_{\a}^{\ \d} (\s_{\un{c}})^{\b\g} ~=~& \frac{1}{16}  \Big\{  \d_{\a}^{\ \b} (\s_{\un{a}\un{b}\un{c}})^{\g\d} + \d_{\a}^{\ \b} \eta_{\un{c}[\un{a}} (\s_{\un{b}]})^{\g\d} \\
    & \qquad  - \frac{1}{2} (\s^{[2]})_{\a}^{\ \b} (\s_{\un{a}\un{b}\un{c}[2]})^{\g\d} - \frac{1}{2} (\s^{[2]})_{\a}^{\ \b} \eta_{\un{c}[\un{a}} (\s_{\un{b}][2]})^{\g\d} \\
    & \qquad  + (\s^{[1]}{}_{\un{c}})_{\a}^{\ \b} (\s_{\un{a}\un{b}[1]})^{\g\d} - (\s^{[1]}{}_{[\un{a}})_{\a}^{\ \b} (\s_{\un{b}]\un{c}[1]})^{\g\d}    \\
    & \qquad  + (\s_{\un{a}\un{b}})_{\a}^{\ \b} (\s_{\un{c}})^{\g\d} - (\s_{\un{c}[\un{a}})_{\a}^{\ \b} (\s_{\un{b}]})^{\g\d} + \eta_{\un{c}[\un{a}} (\s_{\un{b}][1]})_{\a}^{\ \b} (\s^{[1]})^{\g\d}   \\
    & \qquad  + \frac{1}{4!} (\s^{[4]})_{\a}^{\ \b} \eta_{\un{c}[\un{a}} (\s_{\un{b}][4]})^{\g\d} - \frac{1}{3!} (\s^{[3]}{}_{\un{c}})_{\a}^{\ \b} (\s_{\un{a}\un{b}[3]})^{\g\d}     \\
    & \qquad  + \frac{1}{3!} (\s^{[3]}{}_{[\un{a}})_{\a}^{\ \b} (\s_{\un{b}]\un{c}[3]})^{\g\d} - \frac{1}{3!} \eta_{\un{c}[\un{a}} (\s_{\un{b}][3]})_{\a}^{\ \b} (\s^{[3]})^{\g\d}   \\
    & \qquad  - \frac{1}{2} (\s^{[2]}{}_{\un{a}\un{b}})_{\a}^{\ \b} (\s_{\un{c}[2]})^{\g\d} + \frac{1}{2} (\s^{[2]}{}_{\un{c}[\un{a}})_{\a}^{\ \b} (\s_{\un{b}][2]})^{\g\d}  \\
    & \qquad  + \frac{1}{4!3!} \e_{\un{a}\un{b}\un{c}[4][3]} (\s^{[4]})_{\a}^{\ \b} (\s^{[3]})^{\g\d} - (\s_{\un{a}\un{b}\un{c}[1]})_{\a}^{\ \b} (\s^{[1]})^{\g\d}   \Big\}
\end{split}  \\
\begin{split}
    (\s_{\un{b}}^{\ \un{c}})_{\a}^{\ \b} (\s_{\un{c}})^{\d\g} 
    ~=~& \frac{1}{16} \Big\{  -9 \d_{\a}^{\ \g} (\s_{\un{b}})^{\b\d} - \frac{5}{2} (\s^{[2]})_{\a}^{\ \g} (\s_{\un{b}[2]})^{\b\d} - 7 (\s_{\un{b}[1]})_{\a}^{\ \g} (\s^{[1]})^{\b\d}  \\
    & \qquad  - \frac{1}{4!} (\s^{[4]})_{\a}^{\ \g} (\s_{\un{b}[4]})^{\b\d} - \frac{1}{2} (\s_{\un{b}[3]})_{\a}^{\ \g} (\s^{[3]})^{\b\d}  \Big\}
\end{split}  \\
\begin{split}
    (\s_{[\un{a}})^{\a\g} (\s_{\un{b}]}{}^{\un{d}\un{e}})^{\b\d} ~=~& \frac{1}{16} \Big\{  - 2 (\s_{[1]})^{\a\b} (\s_{\un{a}\un{b}}{}^{\un{d}\un{e}[1]})^{\g\d}  \\
    & \qquad  + (\s_{[1]})^{\a\b} \d_{[\un{a}}{}^{[\un{d}} (\s_{\un{b}]}{}^{\un{e}][1]})^{\g\d} - 2 (\s^{[\un{d}})^{\a\b} (\s^{\un{e}]}{}_{\un{a}\un{b}})^{\g\d}  \\ 
    & \qquad  + \d_{[\un{a}}{}^{[\un{d}} (\s_{\un{b}]})^{\a\b} (\s^{\un{e}]})^{\g\d} - \d_{[\un{a}}{}^{[\un{d}} (\s^{\un{e}]})^{\a\b} (\s_{\un{b}]})^{\g\d}   \\
    & \qquad  + \frac{1}{3!} (\s_{[3]})^{\a\b} \d_{[\un{a}}{}^{[\un{d}} (\s_{\un{b}]}{}^{\un{e}][3]})^{\g\d}   \\ 
    & \qquad  + (\s_{[2][\un{a}})^{\a\b} (\s_{\un{b}]}{}^{\un{d}\un{e}[2]})^{\g\d} - (\s^{[2][\un{d}})^{\a\b} (\s^{\un{e}]}{}_{\un{a}\un{b}[2]})^{\g\d}  \\
    & \qquad  - \frac{1}{18} \e_{\un{a}\un{b}}{}^{\un{d}\un{e}[3][\bar{3}]} (\s_{[3]})^{\a\b} (\s_{[\bar{3}]})^{\g\d}  \\
    & \qquad  - 2 (\s_{[1]\un{a}\un{b}})^{\a\b} (\s^{\un{d}\un{e}[1]})^{\g\d} + 2 (\s^{[1]\un{d}\un{e}})^{\a\b} (\s_{\un{a}\un{b}[1]})^{\g\d}  \\
    & \qquad  + \frac{1}{2} \d_{[\un{a}}{}^{[\un{d}} (\s_{\un{b}][2]})^{\a\b} (\s^{\un{e}][2]})^{\g\d} - \frac{1}{2} \d_{[\un{a}}{}^{[\un{d}} (\s^{\un{e}][2]})^{\a\b} (\s_{\un{b}][2]})^{\g\d}  \\
    & \qquad  - \d_{[\un{a}}{}^{[\un{d}} (\s_{\un{b}]}{}^{\un{e}][1]})^{\a\b} (\s_{[1]})^{\g\d} - 2 (\s_{\un{a}\un{b}}{}^{[\un{d}})^{\a\b} (\s^{\un{e}]})^{\g\d}   \\
    & \qquad  - \frac{1}{3} (\s_{[3]\un{a}\un{b}})^{\a\b} (\s^{\un{d}\un{e}[3]})^{\g\d} + \frac{1}{3!} (\s_{[3][\un{a}}{}^{[\un{d}})^{\a\b} (\s_{\un{b}]}{}^{\un{e}][3]})^{\g\d}  \\
    & \qquad  - \frac{1}{3!} \d_{[\un{a}}{}^{[\un{d}} (\s_{\un{b}]}{}^{\un{e}][3]})^{\a\b} (\s_{[3]})^{\g\d} - (\s_{[2]\un{a}\un{b}}{}^{[\un{d}})^{\a\b} (\s^{\un{e}][2]})^{\g\d}  \\ 
    & \qquad  + 2 (\s_{\un{a}\un{b}}{}^{\un{d}\un{e}[1]})^{\a\b} (\s_{[1]})^{\g\d}  \Big\}
\end{split}  \\
\begin{split}
    (\s^{[2]})_{\a}^{\ \b} (\s_{\un{b}[2]})^{\d\g} ~=~& - \frac{9}{2} \d_{\a}^{\ \g} (\s_{\un{b}})^{\b\d} - \frac{1}{2} (\s^{[2]})_{\a}^{\ \g} (\s_{\un{b}[2]})^{\b\d} + \frac{5}{2} (\s_{\un{b}[1]})_{\a}^{\ \g} (\s^{[1]})^{\b\d} \\
    & + \frac{1}{48} (\s^{[4]})_{\a}^{\ \g} (\s_{\un{b}[4]})^{\b\d}
\end{split}  \\
\begin{split}
    (\s^{\un{a}[\un{d}})_{\a}^{\ \g} (\s_{\un{a}}{}^{\un{e}]})_{\b}^{\ \d} ~=~&  \frac{1}{16} \Big\{  12 (\s_{[1]})_{\a\b} (\s^{\un{d}\un{e}[1]})^{\g\d} + 12 (\s^{\un{d}\un{e}[1]})_{\a\b} (\s_{[1]})^{\g\d} \\
    & \qquad  + \frac{2}{3} (\s_{[3]})_{\a\b} (\s^{\un{d}\un{e}[3]})^{\g\d} + \frac{2}{3} (\s^{\un{d}\un{e}[3]})_{\a\b} (\s_{[3]})^{\g\d}  \Big\}
\end{split}  \\
\begin{split}
    \d_{\a}{}^{\b} (\s_{\un{a}})^{\Dot{\g}\Dot{\d}} ~=~& \frac{1}{16} \Big\{  (\s_{\un{a}})_{\a}{}^{\Dot{\g}} C^{\b\Dot{\d}} - (\s^{[1]})_{\a}{}^{\Dot{\g}} (\s_{\un{a}[1]})^{\b\Dot{\d}} - \frac{1}{3!} (\s^{[3]})_{\a}{}^{\Dot{\g}} (\s_{\un{a}[3]})^{\b\Dot{\d}}  \\
    & \qquad  + \frac{1}{2} (\s_{\un{a}[2]})_{\a}{}^{\Dot{\g}} (\s^{[2]})^{\b\Dot{\d}} + \frac{1}{4!} (\s_{\un{a}[4]})_{\a}{}^{\Dot{\g}} (\s^{[4]})^{\b\Dot{\d}} \Big\}
\end{split}  \\
\begin{split}
    \d_{\Dot{\a}}{}^{\Dot{\b}} (\s_{\un{a}})^{\g\d} ~=~& \frac{1}{16} \Big\{  (\s_{\un{a}})_{\Dot{\a}}{}^{\g} C^{\d\Dot{\b}} + (\s^{[1]})_{\Dot{\a}}{}^{\g} (\s_{\un{a}[1]})^{\d\Dot{\b}} - \frac{1}{3!} (\s^{[3]})_{\Dot{\a}}{}^{\g} (\s_{\un{a}[3]})^{\d\Dot{\b}}  \\
    & \qquad  - \frac{1}{2} (\s_{\un{a}[2]})_{\Dot{\a}}{}^{\g} (\s^{[2]})^{\d\Dot{\b}} + \frac{1}{4!} (\s_{\un{a}[4]})_{\Dot{\a}}{}^{\g} (\s^{[4]})^{\d\Dot{\b}} \Big\}
\end{split}  \\
\begin{split}
    (\s^{\un{d}\un{e}})_{\a}{}^{\g} \d_{\Dot{\b}}{}^{\Dot{\d}} ~=~& \frac{1}{16}  \Big\{  - C_{\a\Dot{\b}} (\s^{\un{d}\un{e}})^{\g\Dot{\d}} + (\s^{\un{d}\un{e}})_{\a\Dot{\b}} C^{\g\Dot{\d}} - (\s^{[1] [\un{d}})_{\a\Dot{\b}} (\s^{\un{e}]}{}_{[1]})^{\g\Dot{\d}} - \frac{1}{2} (\s_{[2]})_{\a\Dot{\b}} (\s^{\un{d}\un{e}[2]})^{\g\Dot{\d}}  \\
    & \qquad  + \frac{1}{2} (\s^{\un{d}\un{e}[2]})_{\a\Dot{\b}} (\s_{[2]})^{\g\Dot{\d}} - \frac{1}{3!} (\s^{[3] [\un{d}})_{\a\Dot{\b}} (\s^{\un{e}]}{}_{[3]})^{\g\Dot{\d}} + \frac{1}{4!4!} \epsilon^{\un{d}\un{e}[4][\bar{4}]} (\s_{[4]})_{\a\Dot{\b}} (\s_{[\bar{4}]})^{\g\Dot{\d}}      \Big\}
\end{split} \\
\begin{split}
    (\s^{\un{d}\un{e}})_{\Dot{\b}}{}^{\Dot{\d}} \d_{\a}{}^{\g} ~=~& \frac{1}{16}  \Big\{  C_{\a\Dot{\b}} (\s^{\un{d}\un{e}})^{\g\Dot{\d}} - (\s^{\un{d}\un{e}})_{\a\Dot{\b}} C^{\g\Dot{\d}} - (\s^{[1] [\un{d}})_{\a\Dot{\b}} (\s^{\un{e}]}{}_{[1]})^{\g\Dot{\d}} + \frac{1}{2} (\s_{[2]})_{\a\Dot{\b}} (\s^{\un{d}\un{e}[2]})^{\g\Dot{\d}}  \\
    & \qquad  - \frac{1}{2} (\s^{\un{d}\un{e}[2]})_{\a\Dot{\b}} (\s_{[2]})^{\g\Dot{\d}} - \frac{1}{3!} (\s^{[3] [\un{d}})_{\a\Dot{\b}} (\s^{\un{e}]}{}_{[3]})^{\g\Dot{\d}} - \frac{1}{4!4!} \epsilon^{\un{d}\un{e}[4][\bar{4}]} (\s_{[4]})_{\a\Dot{\b}} (\s_{[\bar{4}]})^{\g\Dot{\d}}      \Big\}
\end{split}
\end{align}
\endgroup

Finally, we list the explicit (real) representations of the sigma matrices in terms of tensor 
products of Pauli matrices:
\begin{align}
(\sigma^{0})_{\alpha\beta} =& \mathbb{I}_{2} \otimes \mathbb{I}_{2} \otimes 
\mathbb{I}_{2} \otimes \mathbb{I}_{2}   \\
(\sigma^{1})_{\alpha\beta} =& \sigma^{2} \otimes \sigma^{2} \otimes \sigma^{2} 
\otimes \sigma^{2}   \\
(\sigma^{2})_{\alpha\beta} =& \sigma^{2} \otimes \sigma^{2} \otimes 
\mathbb{I}_{2} \otimes \sigma^{1}   \\
(\sigma^{3})_{\alpha\beta} =& \sigma^{2} \otimes \sigma^{2} \otimes 
\mathbb{I}_{2} \otimes \sigma^{3}   \\
(\sigma^{4})_{\alpha\beta} =& \sigma^{2} \otimes \sigma^{1} \otimes 
\sigma^{2} \otimes \mathbb{I}_{2}   \\
(\sigma^{5})_{\alpha\beta} =& \sigma^{2} \otimes \sigma^{3} \otimes 
\sigma^{2} \otimes \mathbb{I}_{2}   \\
(\sigma^{6})_{\alpha\beta} =& \sigma^{2} \otimes \mathbb{I}_{2} \otimes 
\sigma^{1} \otimes \sigma^{2}   \\
(\sigma^{7})_{\alpha\beta} =& \sigma^{2} \otimes \mathbb{I}_{2} \otimes 
\sigma^{3} \otimes \sigma^{2}   \\
(\sigma^{8})_{\alpha\beta} =& \sigma^{1} \otimes \mathbb{I}_{2} \otimes 
\mathbb{I}_{2} \otimes \mathbb{I}_{2}   \\
(\sigma^{9})_{\alpha\beta} =& \sigma^{3} \otimes \mathbb{I}_{2} \otimes 
\mathbb{I}_{2} \otimes \mathbb{I}_{2}
\end{align}
and
\begin{align}
(\sigma^{0})^{\alpha\beta} =& -\mathbb{I}_{2} \otimes \mathbb{I}_{2} 
\otimes \mathbb{I}_{2} \otimes \mathbb{I}_{2}   \\
(\sigma^{1})^{\alpha\beta} =& \sigma^{2} \otimes \sigma^{2} \otimes 
\sigma^{2} \otimes \sigma^{2}   \\
(\sigma^{2})^{\alpha\beta} =& \sigma^{2} \otimes \sigma^{2} \otimes 
\mathbb{I}_{2} \otimes \sigma^{1}   \\
(\sigma^{3})^{\alpha\beta} =& \sigma^{2} \otimes \sigma^{2} \otimes 
\mathbb{I}_{2} \otimes \sigma^{3}   \\
(\sigma^{4})^{\alpha\beta} =& \sigma^{2} \otimes \sigma^{1} \otimes 
\sigma^{2} \otimes \mathbb{I}_{2}   \\
(\sigma^{5})^{\alpha\beta} =& \sigma^{2} \otimes \sigma^{3} \otimes 
\sigma^{2} \otimes \mathbb{I}_{2}   \\
(\sigma^{6})^{\alpha\beta} =& \sigma^{2} \otimes \mathbb{I}_{2} \otimes 
\sigma^{1} \otimes \sigma^{2}   \\
(\sigma^{7})^{\alpha\beta} =& \sigma^{2} \otimes \mathbb{I}_{2} \otimes 
\sigma^{3} \otimes \sigma^{2}   \\
(\sigma^{8})^{\alpha\beta} =& \sigma^{1} \otimes \mathbb{I}_{2} \otimes 
\mathbb{I}_{2} \otimes \mathbb{I}_{2}   \\
(\sigma^{9})^{\alpha\beta} =& \sigma^{3} \otimes \mathbb{I}_{2} \otimes 
\mathbb{I}_{2} \otimes \mathbb{I}_{2}
\end{align}

\newpage
$$~~$$

\end{document}